# Interface Engineered Moiré Graphene Superlattices: Breaking the Auger Carrier Multiplication Limit for Infrared Single-Photon Detection

*Sichao Du[1,2#*], Ning Li[1#], Zhufeng Pan[1#], Munir Ali[2,3#], Hengrui Zhang[4], Duokai Chang[1], Yuehang Zhang[1], Qiang Wen[1], Shuo Zhang[3,5], Hao Wu[1,6], Yunlei Sun[1,7], Qiuting Wang[1], Hao Xie[1,2], Chaohao Chen[8], Zhenyi Ni[5], Qiangbing Guo[6], Duo Xiao[1], and Wen-Yan Yin[2*]*

[1]Key Laboratory of Quantum Materials Control of Zhejiang Province, School of Information and Electrical Engineering, Hangzhou City University, Hangzhou, Zhejiang 310015, China

[2] Innovative Institute of Electromagnetic Information and Electronic Integration (EIEI), Zhejiang Province Key Lab of Intelligent Electromagnetic Control and Electronic Integration, State Key Laboratory of Reliability Technology for Electronic Components, College of Information Science and Electronic Engineering, Zhejiang University (ZJU),Hangzhou 310027, China

[3]College of Integrated Circuits, Zhejiang University, Hangzhou, Zhejiang 310058, China

[4]School of Physics, University of Rochester, Rochester, NY 14627, USA

[5]State Key Laboratory of Silicon and Advanced Semiconductor Materials, School of Materials Science and Engineering, Zhejiang University, Hangzhou, Zhejiang 310027, China

[6]State Key Laboratory of Extreme Photonics and Instrumentation, College of Optical Science and Engineering, Zhejiang University, Hangzhou, Zhejiang 310027, China

[7]College of Physics, Zhejiang University, Hangzhou, Zhejiang 310027, China

[8]Faculty of Science, University of Technology Sydney, Ultimo, NSW 2007, Australia

**Corresponding Author\***

Emails: sichaodu@zju.edu.cn; wyyin@zju.edu.cn

#these authors are equally contributed.

**Abstract:**

Two-dimensional (2D) materials are promising candidates for next-generation photodetectors, but their photoelectric conversion efficiency is limited by ultrafast carrier recombination and intrinsic low light absorption. Moiré superlattices have emerged as a versatile platform for engineering optoelectronic properties, yet their potential to overcome fundamental carrier loss mechanisms in high-gain photodetection remains unexplored. Here, we report a 10°-twisted five-layer moiré graphene that breaks the Auger carrier multiplication limit via synergistic engineering of electronic structures and carrier dynamics. The superlattice exhibits additional localized density-of-states at the bilayer interface and enhanced interlayer electron wave coupling, enabling a carrier multiplication gain of ~$10^3$. By engineering a thermalized optical phonon bottleneck, we prolong the hot electron lifetime to ~100 µs, three orders of magnitude longer than conventional 2D materials, thus achieving a maximum Auger scattering rate of ~$10^{15}$ ps$^{-1}$·cm$^{-2}$. Coupling this superlattice with a 90 nm-thick fully depleted silicon-on-insulator (SOI) substrate enables ballistic transport of hot electrons and Schottky barrier-mediated noise blocking. At an ultra-low incident light power of ~$10^{-13}$ W·cm$^{-2}$, the device delivers signal-to-noise ratio exceeding 100 dB and detectivity of $10^{14}$ Jones. Compatible with complementary metal oxide semiconductor (CMOS) processes, the device achieves low-power (60 mW) 100×100 pixel near-infrared imaging at 1310 nm, outperforming mainstream InGaAs/InP and Ge/Si single-photon avalanche diodes (SPADs) in cost, size, and dark count rate. This work demonstrates that moiré superlattice engineering is a viable strategy to tailor non-equilibrium carrier dynamics for high-gain photodetection, bridging nanomaterial science with electronic engineering for practical infrared sensing applications.

# Introduction

Infrared photodetectors are critical components in autonomous driving, deep-space exploration and biomedical imaging, with growing demand for high gain, low power consumption and CMOS compatibility.[1, 2] 2D materials offer sensitive photodetection and ultrafast carrier transport, but their performance is hindered by two key limitations: the short lifetime of charge carriers (rapid recombination of photo-generated charge-carriers, ~100 fs) and the inherently low light absorption of two-dimensional materials (due to their ultrathin nature, ~ 3 nm).[3, 4]

Existing strategies to address these issues including heterostructure design, waveguide integration and nanoparticle coupling etc., focus on external structural optimization without fundamentally modifying the carrier multiplication mechanism. [5-13]. As a result, the carrier multiplication gain in 2D material-based photodetectors has long been stuck below 5.[14-17] Moiré superlattices, formed by twisting two or more 2D layers to create periodic atomic arrangements, have emerged as a powerful platform to engineer electronic properties via twist-angle-controlled interlayer interactions. Previous studies have shown that moiré graphene superlattices exhibit tunable band structures and enhanced electronic correlations, but their application in photodetection with high carrier multiplication gain remains unexplored.

Here, we report a ballistic carrier multiplication barristor based on a 10°-twisted five-layer moiré graphene superlattice integrated with a fully depleted SOI substrate. The device leverages two intrinsic effects of the moiré superlattice: enhanced localized density-of-states and interlayer electron wave coupling, to boost Auger scattering efficiency. By engineering a thermalized optical phonon bottleneck, we extend the hot electron lifetime to ~100 μs, enabling efficient carrier multiplication. The integration with SOI further amplifies the gain via ballistic avalanche and suppresses noise through Schottky barrier blocking, resulting in a total gain of ~$10^7$.

# 1. Device Working Principle

The carrier multiplication transport process in the graphene moiré superlattice and the device energy band structure in this work are illustrated in Figure 1a. The main working mechanism includes Auger scattering, interlayer coupling, Relaxation bottleneck, ballistic avalanche and noise blocking. First, the enhanced light absorption brought by the moiré graphene superlattice structure can produce more hot electrons with energy higher than the Fermi surface. The twisting angle-controlled electron wave function enhances the interlayer coupling and increases the scattering among the hot electrons. Secondly, the increase of electron correlation states at moiré graphene interlayer interface enhances the efficiency of Auger scattering among electrons. Hence, the hot electrons can multiply more cold electrons per unit time. Because electrons are concentrated in a narrow energy range, resulting in extremely high electron density-of-states and accelerating hot electron scattering.[18] Thirdly, by tuning the top and bottom gate voltages, the competitive cooling path of optical phonon (thermalized phonon effect[12, 19]) or acoustic phonon (super-collision[20, 21]) can be adjusted in the hot electrons' relaxation process. Therefore, the relaxation "bottleneck" can be established to achieve the maximum Auger carrier multiplication in the time domain. Fourth, the lateral dimension of silicon-on-insulator (SOI) silicon is optimized so that its thickness is less than the average free path of hot electrons. Then the hot carriers ballistic avalanche can be realized. Fifth, the Schottky barrier height effectively blocks the white noise injected from the moiré superlattice.

The photo-current time response and corresponding current-voltage curves under different bias voltages prove what we call the carrier multiplication process, as shown in Figure 1b-d. The integral of each curve as a function of time is equal to the number of multiplied hot electron collections. Although the micro-seconds time scale is mainly due to the capacitance of the depleted SOI silicon junction, the accumulation time of hot electrons is extended by nearly 100 μs when the top gate voltage is higher than 10 V. This means that our device has achieved extremely high hot electron multiplication gain (SI Figure 1-2). The essential reason is that the generated hot electrons are not rapidly relaxed by acoustic phonons. We design optical phonons as the main relaxation path by optimizing the moiré superlattice thickness, SOI silicon substrate thickness, incident laser fluence and gate voltages to utilize the hot phonon effect.[22] When there are a large number of scattered high-energy hot electrons in the system, the energy dissipation rate reaches "bottleneck"[23, 24]. It makes the hot electrons forced to stay in the electronic system and continue to undergo Auger carrier multiplication. Finally, the excess energy in the system is gradually quenched by acoustic phonons and dissipated in the form of Joule heat of the SOI silicon substrate, thus making the whole system in equilibrium. The corresponding mechanism flow as a function of time is displayed in Figure 1c. The extremely high gain due to the increase in the number of hot electrons inevitably leads to the enhancement of photo-current, which means that the efficiency of converting incident optical power into photo-current is improved. Our device achieves $10^7$ gain with $10^4$ A/W and $10^{14}$ Jones at $10^{-4}$ mW/cm$^2$, as displayed in Figure 1e and f.

## 2. Device and Materials Structures

### 2.1. Device Structure

We investigate the ultrafast energy relaxation dynamics in moiré graphene superlattice to explore the intrinsic mechanisms behind its carrier multiplication. As depicted in Figure 2a, a four-terminal ballistic carrier multiplication barristor (BCMB) device structure is proposed, where the cascade avalanche multiplication of hot electrons plays a critical role in achieving efficient photodetection. Top metallic gating and bottom optoelectrical gating are used to modulate the hot carrier dynamics within the moiré superlattice. The multiplication gain provided by the hot carrier relaxation dynamics in the moiré superlattice is further amplified through ballistic transport of tunneled carriers undergoing cascaded avalanche in the thin, deeply depleted SOI silicon layer. Further device fabrication information can be found at SI Figure 3-4.

### 2.2. Materials Structure

Figure 2b is the visible-near infrared absorptivity spectrum of the moiré graphene superlattice we used in the work. Under the condition of a constant twist angle of 10º between each two moiré graphene layers, the absorption peak intensity gradually increases as the number of graphene layer increases from 2 to 5. When the number of graphene layer is 5, the absorptivity reaches 5.5%. The absorptivity peaks appear near 950 nm, which corresponds to the energy position of the additional interlayer localized density-of-states of a 10º bi-layer moiré graphene.[25] This phenomenon suggests that due to the synergistic enhancement of interlayer interactions, the twisted graphene exhibits an interlayer coupling effect that is different from the previous few-layer structures.[26-28] Moreover, the step height profiles between the etched silicon and the surrounding SiO$_2$ show that the top silicon layer of the SOI was thinned to create a fully depleted semiconductor layer, ensuring ballistic avalanche[29], as depicted in Figure 2c. Furthermore, the SOI substrate provides full dielectric isolation between adjacent devices, leading to lower leakage currents and reduced capacitive coupling during avalanche voltage operation. Figure 2 d-f shows the angle-resolved photoemission spectroscopy (ARPES) spectrum of the moiré graphene superlattice. The ARPES measurements suggest that four Dirac-type dispersions near the K point are twisted with a moiré angle of 10º between each other. Further materials characterizations can be found at SI Figure 5-6.

**2.3. Device Performance**

As the core index to evaluate performance of photodetectors, a maximum of signal-to-noise ratio needs to be achieved through a synergistic optimization of "enhancing effective signal" and "suppressing useless noise".[30] Our work achieved an ultra-high signal-to-noise ratio of ~100 dB based on the multi-dimensional control of the whole charge carriers transport chain. In terms of noise interception, the thickness of SOI silicon is optimized to realize the ballistic transport of hot electrons as depicted in Figure 2g and SI Figure 7d. Hence, the noise generated by carrier scattering is minimized. The small thickness of moiré superlattice employs that a strong electrical field exceeding 30 kV/cm$^2$ is generated throughout the moiré graphene layer with a bias voltage of 11 V. Complete electrical isolation of the moiré graphene layer from the SOI silicon also blocks the unwanted slow diffusion of photo-generated carriers away from the high field region. Moreover, the Schottky barrier is used to block the white noise injected from moiré graphene superlattice as evidenced in Figure 2h. At the signal gain level, the optimal path of hot electron relaxation (balance of optical phonon and acoustic phonon cooling[31]) is tuned by the gating voltage. This strengthens the Auger carrier multiplication and the thermalized optical phonon relaxation bottleneck effect. It then enhances the hot electrons accumulation to achieve high gain and amplify the photo-current signal. This is also confirmed by temperature-dependent current-voltage swipes, which shows a periodical oscillation of the Fabry–Perot type of interferences under 100 K (SI Figure 8). Furthermore, the temperature-dependent diode characteristic fittings depict the ballistic transport regime in contrast to the conventional diffusion transport regime, as depicted in Figure 2i. The plot of the natural logarithm of the saturation current ($I_{sat}$) multiplied by the square of the temperature ($T^2$) versus the inverse temperature ($1/T$). The linear trend suggests that the $I_{sat}$ follows an Arrhenius-type behavior, which is representative for thermally activated charge carriers transport. The positive slope indicates an increase in $I_{sat}$ with

T, which is related to the enhanced carrier generation and a decrease in energy barriers for charge carriers transport.

## 3. Hot Electrons Dynamics

### 3.1. Transient Absorption Measurements

Then, we study the dynamics of photo-excited hot Dirac Fermions in the moiré graphene superlattice by using the transient absorption spectroscopy under a wide photo-excitation energy range, as displayed in SI Figure 9. The work function has been confirmed as an almost intrinsic doping statistic as ± 20 meV. Since the pump photon energies are above twice the value of the Fermi energy of the moiré graphene sample, no intra-band transitions initially occur. All the photo-excitation energies result in a photo-bleaching for the differential transmissivity $\Delta T/T_0$. This represents that the hot electrons' occupation prevents the filling of electrons excited by the probe pulse due to the Pauli blocking. Hence, these contour maps indicate the impact ionization induced inter-band scattering process, leading to a carrier multiplication effect as illustrated in Figure 3a.

These inter-band transitions lead to different hot electron scattering dynamics monitored in the probe energy range from 400 to 450 meV, as shown in SI Figure 10. The $\Delta T/T_0$ signals illustrated in Figure 3b-c correspond to the hot electrons in the probe energy range lower than the initial photo-excitation energy state. While the $\Delta T/T_0$ signals illustrated in Figure 3e-f correspond to the hot electrons in the probe energy range higher than the initial photo-excitation energy state. This is a clear signature of hot electrons scattered from the lower energy state, indicating the Auger recombination induced intra-band scattering process as illustrated in Figure 3d. In this colour scale, the change of the $\Delta T/T_0$ amplitude is within three orders of magnitude. Higher photo-excitation energy implies a larger number of generated hot electrons. This also confirms that the strength of the Auger carrier multiplication scattering increases as sufficient photo-excitation energy is offered, as shown in SI Figure 11. The $\Delta T/T_0$ displayed represents the multiplied hot electrons, scattered with cold electrons from the valence band through impact ionization.

### 3.2. Carrier Multiplication Analysis

Then, we explicitly calculate the carrier multiplication (CM) factor, which is given by the ratio of the generated electrons density $N_G(\tau, \varepsilon)$ and the optically injected carrier density $N_I(\tau,\varepsilon)$, $CM = N_G/N_I$. The total excited carrier density $N_G$ remains above the initially optically injected density $N_I$. The integrated carrier density is extracted by numerical integration $N_G(\tau) = \int n_{(E,\tau)}dE$ of the fitted population curves as a function of time delay $\tau$. An efficient CM requires a long time range with a predominant impact excitation. The generation rate of the hot electrons then can be calculated (Figure 3g), with the peak value delayed as the pump photon wavelength increases from 905 to 1550 nm. However, as the pump photon wavelength increases, the peak value of the population density decreases as depicted in Figure 3h. These calculations directly investigate the appearance of CM as a function of pump-probe delay time within the pump fluence regime experimentally analyzed. The higher the fluence, the greater the scattering, and the more efficient the initially relevant impact excitation, resulting in a larger $N_G$. Figure 3i show the CM as a function of the pump photon wavelength at 300 fs, 500 fs, 800 fs and 1 ps. The CM decreases as the pump photon wavelength increases. It increases from 300 fs to 500 fs, then decreases from 500 fs to 1 ps. After 2 ps, the CM approaches zero.

The strong electronic correlations from the moiré graphene superlattice give rise to an enhanced photo-sensitivity of our device, while the high electric field strength ensures fast acceleration of multiplied electrons to their saturation velocities in a ballistic avalanche regime. In order to enhance the carrier multiplication effect of graphene, we need to provide more phase space for it. In view of the enhanced strong electron correlation energy state at the moiré graphene interface, we utilize the energy band flattening effect of the moiré graphene. This greatly increases the local density-of-states and provides more phase space for carrier multiplication, thus promoting electron scattering under the Auger effect (SI Figure 10 e-f). The weakly coupled twisted graphene moiré superlattice structure produces a cascade effect and capture more hot carriers for the subsequent ballistic avalanche.

## 4. Imaging Application

The mainstream technology route of Light Laser Detection and Ranging (LiDAR) is currently based on the InGaAs/InP[32] and Ge/Si[33] single-photon avalanche-diode (SPAD) array. The InGaAs/InP SPAD normally suffers from high cost, large volume and complexity of the fabrication. While for the Ge/Si SPADs, a 4.2% lattice mismatch at the Ge/Si interface introduces defects, resulting in an increase in dark count rate (DCR) of the pixels. This requires the addition of buffer layers and multi-step passivation compensation. In contrast, our carrier multiplication barristor is a Schottky-type, shallow-junction avalanche photodetector fabricated using a complementary metal oxide semiconductor (CMOS) process. The top and bottom layers of moiré graphene superlattice serve as protective barriers, while the inner layers of moiré graphene superlattice remain intrinsic. By generating dynamical gate electric field, the photo-induced carrier multiplication allows the device to benefit from the noise reduction, which arises from clean interfaces among moiré graphene layers. These results offer a practical method for enhancing the performance of single-photon avalanche detectors, silicon photomultipliers, and image sensors.

Figure 4a illustrates a schematic of the single-pixel scan-imaging approach we used in this work with a mask of a Chinese character "杭 (Hang)". Figures 4 b-c show the photon-sensing window of our device with corresponding two-dimensional photo-current mapping under a 1310 nm pulsed laser. Extension of three-dimensional imaging into the infrared range gives less interference from ambient background visible light. Hybrid integration of moiré graphene superlattice photo-transistors with high-density pixels has been demonstrated in a 100 × 100 scanning image array. Figure 4d presents the room-temperature infrared images we captured by the time-correlated 1310 nm pulsed laser with corresponding photo-responsivity grey scale.

The conversion of light into free electron-hole pairs constitutes the key process in the fields of photodetection.[34, 35] The efficiency of this process depends on the competition of different relaxation pathways. It can be greatly enhanced when photo-excited carriers do not lose energy as heat, but instead transfer their excess energy into the production of additional electron-hole pairs through carrier multiplication effect. Our results indicate that carrier scattering is highly efficient, prevailing over phonon scattering and leading to the production of secondary hot electrons originating from the valence band. From the analysis of transient absorption spectroscopy, the carrier multiplication gain from the moiré superlattice yields a value of ~$10^3$. The avalanche gain from the deep depleted silicon is ~$10^4$, resulting in an overall gain of ~$10^7$ under the photon flux in an order of $10^9$ photons/cm$^2$. Therefore, the photo-sensitivity of our device as a function of the irradiance shows a trend approaching to a single-photon detection limit as shown in Figure 4e. Then we can calculate to estimate the power consumption of our SPAD scanning array, where we have a 100 ×100 pixels, and $V_{bias}$ = 2.8 V. Since $I_{static}$ = 1 pA and the $I_{static,total}$ = 100 × 100 × 1 pA = 10 nA, the $P_{static}$ =

2.8 V × 10 nA = 28 nW, as depcited in Figure 4e. If the average trigger rate per pixel is 1 KHz, $E_{avalanche}$ = 1 nJ. Then $P_{avalanche}$ = 100 × 100 × 1000 × 1 nJ/s = 10 mW. The power consumption of the read-out circuit is about 50 mW. Therefore, $P_{total} \approx P_{static} + P_{avalanche} + P_{read-out}$ = 60 mW, enables a low power consumption of our active device sources as depicted in Figure 4e.

## Conclusion

Van der Waals atomic layers can be designed as artificial nanoelectronic building blocks through the deterministic assembly of multiple layers. These Van der Waals crystals possess weak interlayer interactions, which allow arbitrary azimuthal orientations between the 2D lattices to be independently controlled. The twisting-angle-dependent electronic correlations lead to efficient carrier multiplication, particularly in the context of charge carriers harvesting. Our high gain is attributed to the hot electrons' energy relaxation bottleneck which is induced by the thermlized optical phonons. This effect greatly enhances hot electrons' accumulation to prolong the Auger carrier multiplication process. Notably, the band structure of moiré superlattice is highly sensitive to the applied perpendicular gating electrical field, thus providing an additional degree of control over the relative strength of electronic correlations. This is further supported by coupling the moiré superlattice to a deeply depleted thin silicon film from a SOI platform, which offers CMOS compatibility for large-scale fabrication. The thin silicon film additionally facilitates conventional impact ionization-based gain for carriers that tunnel through the $SiO_2$ layer from the moiré superlattice. We anticipate that the engineering of moiré superlattices will become a general strategy for tuning carrier multiplication across a range of 2D materials.

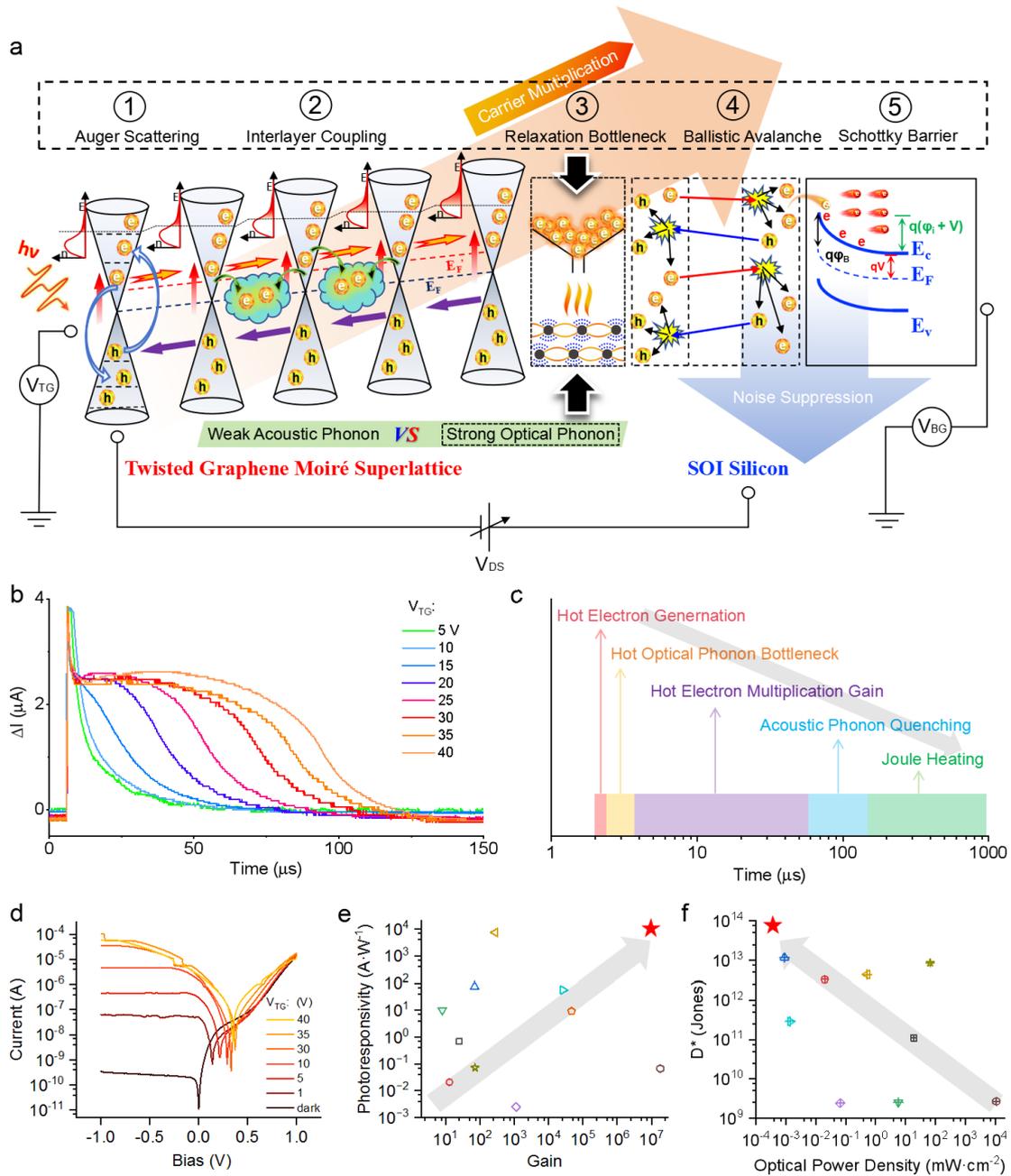

**Figure 1**. **The Working Principle and Device Performance. a**, The schematic of energy band diagram and the charge carriers transport of the device. **b**, a single time response cycle of the device under different gate voltages. **c**, the dominant effects in a single time response cycle of the device. **d**, the current-voltage curves of the device under different gate voltages. **e**, the relationship between the photoresponsivity and gain of different types of devices in the literature, where the red star represents the parameter of this work. **f**, the relationship between the detectivity as a function of optical power densities of different types of devices in the literature, where the red star represents the parameter of this work. Further information can be found at SI Table I.

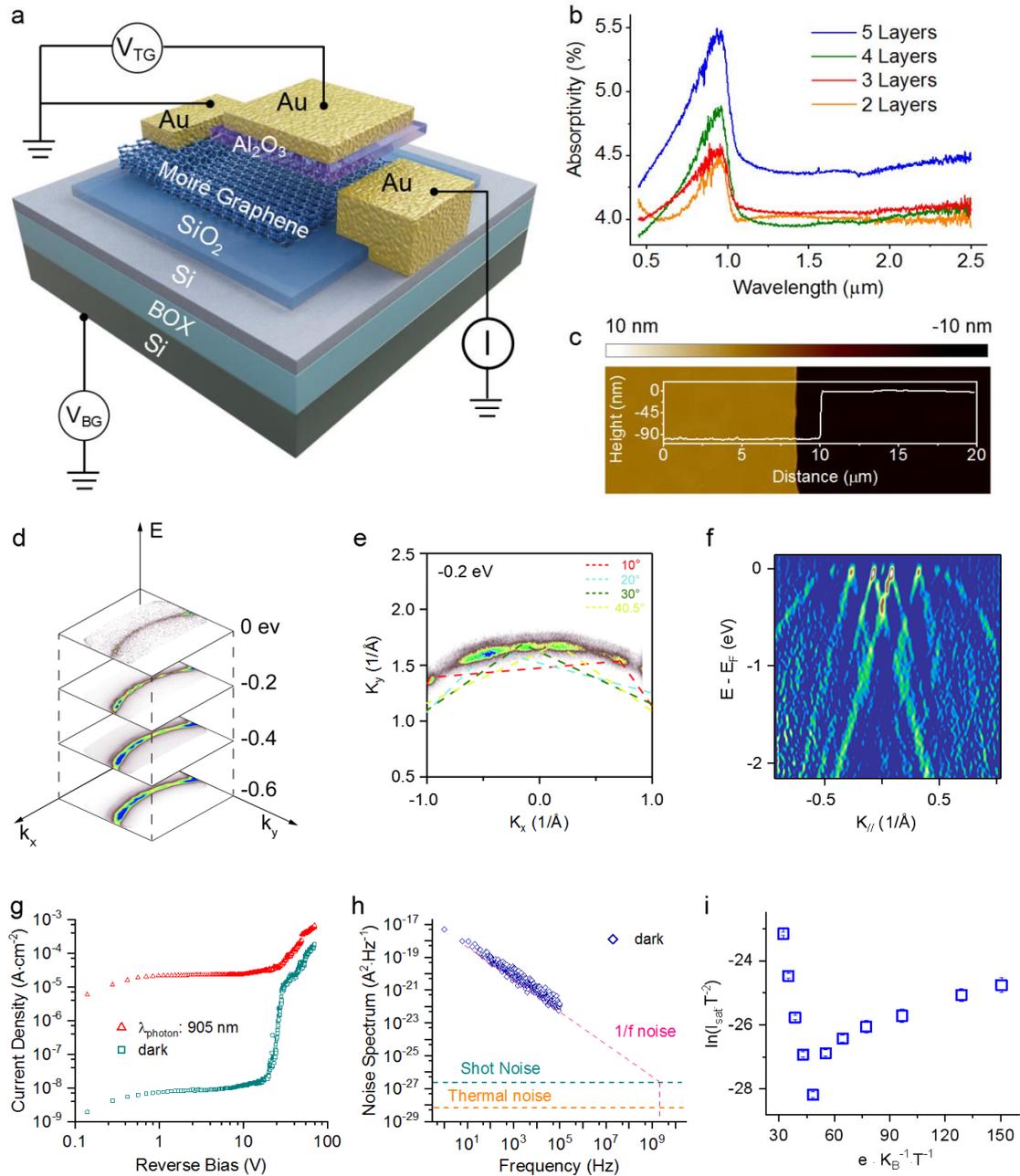

**Figure 2**. **The Device and Materials Structures. a**, The schematic device structure. **b**, the absorptivity of the moiré graphene with different graphene layer numbers as a function of the light wavelength. **c**, The top view of AFM mapping and height profile of the step between silicon and silicon oxide. **d**, The two-dimensional curvature image of constant energy maps of the ARPES characterization of moiré graphene. **e**, at the energy of $-0.2$ eV. **f**, The second convolutions of the measured Dirac-type dispersion near the K point of the corresponding orientation. **g**, The current density as a function of the reverse bias of the device under dark and light condition. **h,** The noise spectrum as a function of the scanning frequency of the device. **i**, The measurement temperature-dependent diode characteristic fittings.

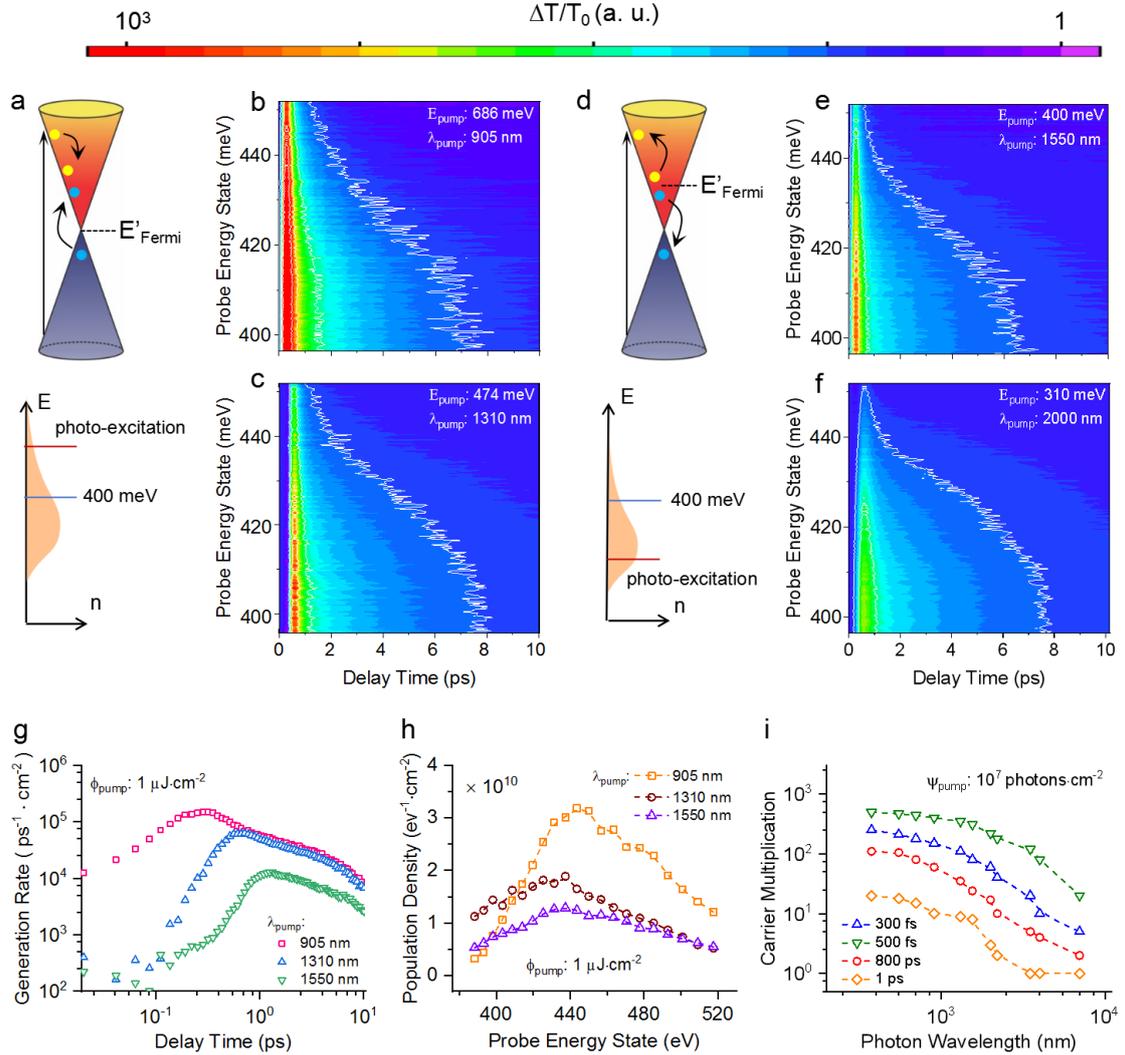

**Figure 3. The Transient Absorption Characterization of Carrier Multiplication.** The schematic of the impact ionization induced **a,** carrier multiplication and **d,** Auger recombination with corresponding Fermi-Dirac distributions in a single-layer graphene. The carrier density *n* as a function of the energy state *E* depicts the pump photo-excitation energy height and the bottom position of the probe energy at 400 meV. The dynamical quasi-Fermi level $E'_{Fermi}$ is also shown to distinguish the inter- and intra- band transitions. The two-dimensional differential transmissivity countor maps as functions of the probe energy state and delay time of the moiré graphene under pump photon energy $E_{pump}$ (and corresponding pump photon wavelength $\lambda_{pump}$) of **b**, 686 meV (905 nm); **c,** 474 meV (1310 nm); **e,** 400 meV (1550 nm) and **f,** 310 meV (2000 nm). **g,** The hot electron generation rate as a function of the delay time. **h,** The hot electron population density as a function of the probe energy state. **i,** The extracted carrier multiplication as a function of the pump photon wavelength at the delay times 300 fs, 500 fs, 800 fs and 1 ps.

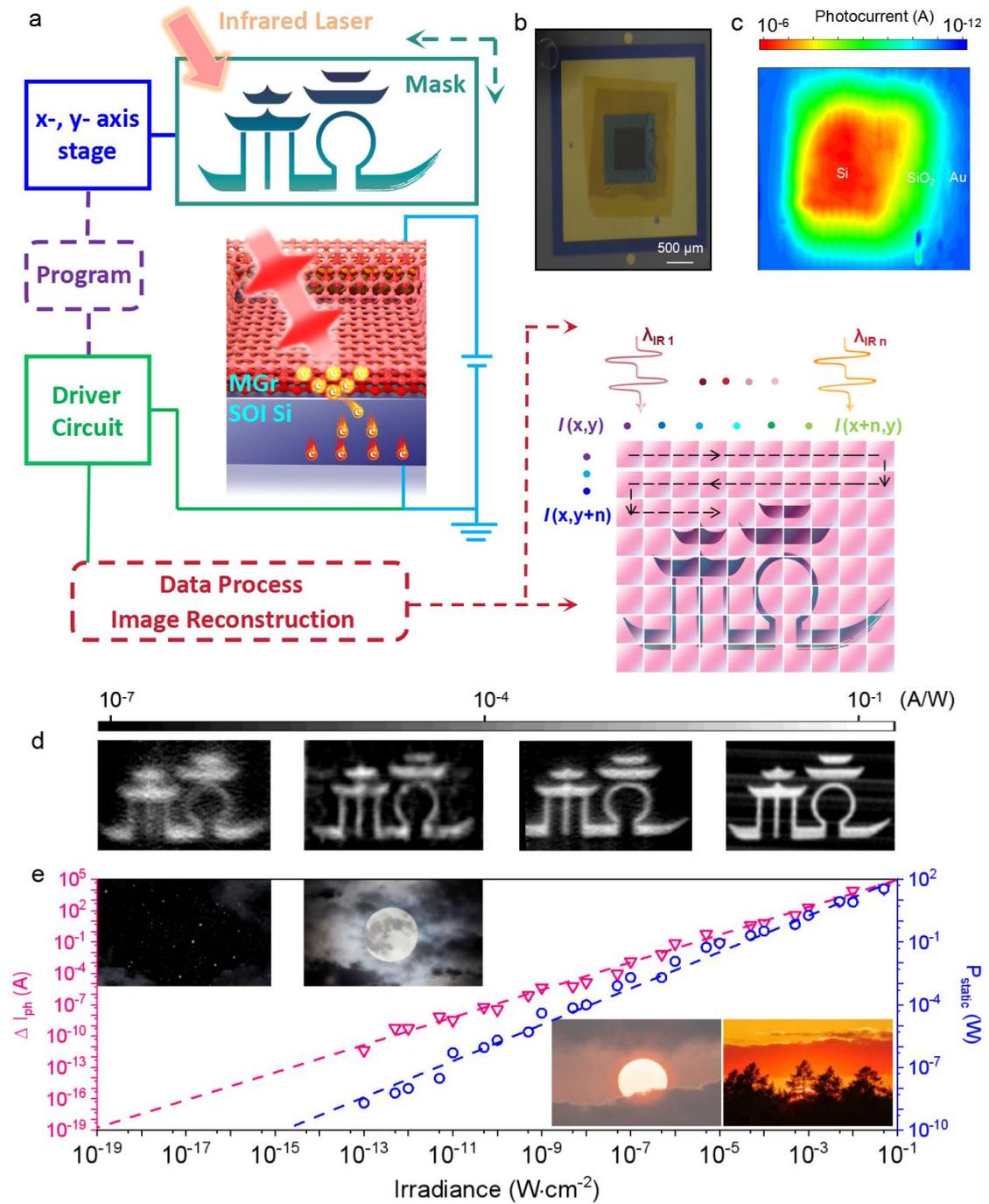

**Figure 4. The Image Sensing Characteristerizations**. **a**, The schematic of the single pixel imaging principle in this work. **b**, The photon sensing window of the the single pixel device, and **c**, the corresponding photo-current mapping under a 1310 nm laser pulse with fluence of a 10 nJ·cm$^{-2}$. **d**, The corresponding grey-scale scanned images. **e**, the photo-current of the single

pixel and the static power consumption of the scanned array as a function of the irradiance, where the environmental photos indicate corresponding irradiance values.

## Methods

**The Device Fabrication**

The SOI wafers with a 10 μm thick, lightly-doped (1–10 Ω cm) n-type silicon device layer (100) were employed to fabricate a planar moiré graphene silicon Schottky diode. The top silicon was thinned down to 25 to 200 nm by wet etching of silicon in in 30 wt% KOH solution at 80 °C and subsequent dry etching (inductively coupled plasma reactive ion etching, ICP-RIE). The top silicon was coated with a 100 nm thick silicon-dioxide ($SiO_2$) layer by inductively coupled plasma enhanced chemical vapor deposition (ICP-PECVD, Oxford Instruments). Subsequently, two areas are opened in the $SiO_2$ layer for the formation of the junction and the metal-silicon contact. The size of silicon windows are from 50 μm × 50 μm to 500 μm × 500 μm. These areas are defined by optical lithography. The oxide layer in these areas is removed by wet etching in a buffer-oxide-etch (BOE) solution. Then contacts to the top silicon layer are fabricate by an additional lithography step followed by thermal evaporation of gold/chromium (5/100 nm) and lift-off. A metal frame to electrically contact graphene has been designed around the moiré graphene–silicon junction of the planar device to minimize series resistance. A dielectric separation with the top gate is achieved with 10 nm of $SiN_x$ deposited on the silicon surface by ICP-PECVD (Oxford Instruments), followed by $Al_2O_3$ atomic layer deposition. The top gate of chromium/gold (5 nm/50 nm) is deposited along the center of the silicon window. The devices were annealed for 120 minutes at 300 ºC under $H_2$ before all the measurements to remove the residues and contaminants. Au wires were then bonded with the electrodes.

**The Device Measurements**

A Fabry–Perot quantum cascade laser was used as the light source, which was chopped to a specific frequency of light. The laser beams were focused by a reflection objective. The photo-voltage responses were measured using the lock-in and current amplifier, Stanford SR830 and SR560, respectively. The photocurrent responses were measured using the agilent B1500A and Keithley 2602B dual-channel digital source meter. The noise was measured by a noise measurement system (PDA NC300L, 100 kHz bandwidth). The device and the trans-impedance amplifier were connected with an oscilloscope (Keysight UXR1102B, 110 GHz bandwidth) to measure the photo-voltages. Infrared bandpass filters (FB2000-500 to FB4000-500, Thorlabs) were used to filter out the stray light from the infrared laser. Neutral density (ND) filters were used to vary the light intensity. A calcium fluoride aspherical lens system was used to focus the infrared light on the device. The power was measured by a photodiode and thermopile detector (Newport 818/918). The temperature dependent menasurements were conducted in the Lakeshore CPX low temperature probe station. For the photograph, we built a scanning systemwith a focused laser and a programmable X-Y stage. A data acquisition (DAQ) card was used to obtain the photocurrent data. The measurement results for each sample were recorded in identical conditions to ensure comparable results. The samples were tested a number of times, showing no damage and repeatable measurement results.

**The Pump Probe Measurements**

For femtosecond TA and TR spectroscopy (Femto-TR100, Time-Tech Spectra LLC), the fundamental output (1030 nm, ~190-fs pulse duration, 200 µJ/pulse, and 100 kHz) from a Yb: KGW laser (PHAROS, Light Conversion Ltd.) was separated into multiple light beams. One was introduced to an optical parametric amplifier (OPA, Orpheus-HP, Light Conversion Ltd.) equipped with difference frequency generation (DFG) to generate near-IR wavelengths as the probe beam. The other was directed onto an yttrium-aluminum-garnet (YAG) crystal to generate a white light continuum as the pump beam.

**The Materials Preparation**

The moiré graphene samples were grown on a nickel foil via the CVD method.[36, 37] A number of graphene samples with different thicknesses were grown. The moiré-graphene-superlattice/nickel film was spin-coated by poly-methylmethacrylate (PMMA) with a thickness less than 100 µm. The spin speed was 500 rpm for the first

5s and 4000 rpm for the next 60s. Next, the PMMA/moiré-graphene-superlattice/nickel sample was placed in a nickel etching solution for 12 h to etch the nickel substrate. The etched film was then cleaned with de-ionized water. Thereafter, PMMA/moiré-graphene-superlattice was transferred onto the SOI silicon substrate. Finally, the PMMA was washed off using dichloromethane at 50 ºC and isopropyl alcohol.

**The Materials Characterization**

The thickness of moiré graphene was initially identified by an optical microscope and finally determined by an atomic force microscope (Bruker Multi-Mode 8). Scanning electron microscopy (SEM) images were taken on a Hitachi S4800 field emission system at an acceleration voltage of 3 kV. Raman spectroscopy was performed using a confocal Raman microscope (Senterra, BRUKER) and a Renishaw Invia system with a 532 nm laser. Wide-angle X-ray scattering measurements and angle-resolved photoemission spectroscopy were carried out at the Shanghai Synchrotron Radiation Facility in Shanghai, China. The energy of the X-rays was 18.981 keV and the wavelength was 0.653 Å.

## Acknowledgements

The authors acknowledge the facility support by The State Key Laboratory of Molecular Reaction Dynamics, Dalian Institute of Chemical Physics; Shanghai Institute of Optics and Fine Mechanics; National Laboratory for Infrared Physics, Shanghai Institute of Technical Physics, Chinese Academy of Sciences. The State Key Laboratory of Superlattices and Microstructures, Institute of Semiconductors. National Laboratory of Solid State Microstructures, School of Electronic Science and Engineering, Nanjing University, Department of Chemistry, Zhejiang University. The authors also would like to acknowledge the material contributions from Rodney S. Ruoff in Ulsan National Institute of Science and Technology. The authors also would like to thank the help from Dr. Peng Li, Dr. Srikrishna Chanakya Bodepudi, Dr. Muhammad Abid, Dr. Muhammad Malik, Dr. Nasir Ali, Dr. Peng Wang, Dr. Fang Zhong, Dr. Qiaohui Zhou, Prof. Haiming Zhu, Dr. Runchen Lai, Dr. Junxue Liu, Prof. Kaifeng Wu, Prof. Shengye Jin, Dr. Yujie Peng, Dr. Chunhui Zhu, and Prof. Fengqiu Wang.

# Funding

This work was supported by National Natural Science Foundation of China (Grant No. 62474155 and 61704150), Natural Science Foundation of Hangzhou (Grant No. 2024SZRZDA040001), High Level Returned Overseas Scholars in Hangzhou (Grant No. 202000-591807), Young Talents Plan of Zhejiang University City College (Grant No.202000-581832).

# Conflict of Interests

The authors declare no conflict of interest.

# Data Availability

The data that support the findings of this study are available from the corresponding authors upon reasonable request.

# Supplementary Section 1

# Avalanche breakdown characteristics

## A.     Ballistic avalanche transport phenomenon

The temperature dependent I-V scan images confirmed the carrier ballistic avalanche characteristics of the device (SI Figure 7-8). The scanning shows the periodic oscillation of Fabry Perot type interference at 100k, which is caused by the quantum interference of the multiple reflection path of the carrier between the two partially reflecting electrodes. The bias controls the kinetic energy of carriers, thus adjusting their wavelength. The properties of this quantum mechanical wave clearly show that the propagation of carrier wave in SOI silicon is ballistic - there is no scattering to a large extent. Compared with the traditional diffusion transport mechanism, the temperature dependent diode characteristics determine the ballistic transport mechanism.

The noise spectrum is divided into three parts: (1) thermal noise or Johnson noise, (2) shot noise, and (3) 1/f noise. When it is less than 100kHz, it shows a classic 1/f shape consistent with the pink inclined dotted line, which is different from the excessive noise (white noise) of traditional APD. The green horizontal dotted line represents the theoretical limit of traditional shot noise, and the yellow horizontal dotted line represents the theoretical limit of traditional thermal noise. The spectral density of thermal noise is given by Nyquist formula,

$$S_I(f) = 4k_B T/R$$

Where $k_B$ is the Boltzmann constant and t is the temperature. The spectral density of shot noise is given by Schottky theorem,

$$S_I(f) = 2q <I>$$

Where $<I>$ is the average value of current. Thermal noise and shot noise are caused by the random motion of charge carriers. These two kinds of noise are called white noise because their spectral density does not depend on $f$.

Different from other types of intrinsic noise, 1/f noise mainly comes from carrier density and mobility, and is dominant at low frequencies. The most commonly used quality factor - Hooge parameter, $\alpha_H$ - is based on its empirical formula,

$$S_R/R^2 = \alpha_H/Nf$$

Where $S_R \sim (\delta R)^2$ is the power spectral density of resistance fluctuation $S_R/R^2 = S_I/I^2 = S_V/V^2$, and V is the voltage. The sensitivity and selectivity of many types of sensors, especially those that rely on electrical response, are limited by 1/f noise. Although 1/f noise is dominant only at low frequencies, it will be upconverted to high frequencies due to the inevitable nonlinearity in devices or systems. Therefore, 1/f noise is the main source of phase noise in communication systems and sensors. Reducing the size of any material system for nanoscale devices will further increase the 1/f noise level and complicate the practical application.

## B. Carrier concentration in silicon

For intrinsic semiconductors, the carrier concentration can be calculated by the following formula,

$$n_e n_h = 4\left(\frac{k_0}{2\pi\hbar^2}\right)^3 (m_e^* m_h^*)^{3/2} T^3 \exp\left(-\frac{E_g}{k_0 T}\right)$$

Where $n_e$ is the electron concentration, $n_h$ is the hole concentration, $k_0$ is the Boltzmann constant, $\hbar$ is the reduced Planck constant, $m_e^*$ is the effective mass of the electron, $m_h^*$ is the effective mass of the hole, $T$ is the temperature, and $E_g$ is the band gap. For the effective mass $m^*$ of the electron or hole, it can be calculated according to the following formula,

$$\frac{1}{m^*} = \frac{1}{\hbar^2}\frac{\partial^2 E}{\partial k^2}$$

Where $k$ is the wave vector and $E$ is the energy. The effective mass used in the mobility is the transport effective mass, and the longitudinal and transverse effective masses of electrons or holes need to be statistically averaged, $m_e^* \approx (m_l \cdot m_t^2)^{1/3} \approx (0.98m_0 \cdot 0.19m_0^2)^{1/3} \approx 0.26m_0$, $m_l$ is the longitudinal effective mass and $m_t$ is the transverse effective mass. The effective mass can also be measured directly through the cyclotron resonance experiment. Take silicon intrinsic semiconductor as an example, $m_e^* = 0.26m_0$, $m_h^* = 0.39m_0$, $m_0 = 9.109 \times 10^{-31} Kg$ is the electronic static mass, the band gap $E_g$ of silicon is 1.12eV. According to the formula, $n_e = n_h \approx 1.15 \times 10^{10} cm^{-3}$.

## C. Average free path and mobility

Lattice scattering results from the thermal motion of lattice atoms above absolute zero. The lattice vibration increases with the increase of temperature. The average free path at high temperature is dominated by lattice scattering, and its average free path path can be expressed by the following formula,

$$l = \frac{3\sqrt{2\pi m^* k_0 T}}{4e}\mu$$

Where $l$ is the average free path, $\mu$ is the mobility, and $m^*$ is the effective mass of the electron or hole. The mobility can be calculated by the following formula,

$$\mu = \frac{J \cdot L}{neV_{ds}}$$

$J$ is the current density, $n$ is the carrier concentration, and $L$ is the thickness of the material. Electron mobility in intrinsic silicon $\mu \approx 2000 cm^2/(V \cdot s)$, according to the above formula, $l \approx 74\ nm$. The average free path at room temperature can also be estimated by thermal velocity, $l = V_{th} \cdot \tau$, $\tau$ is the average free time, $V_{th}$ is the average thermal velocity, and $k$ is the Boltzmann constant,

$$\frac{1}{2}m^*V_{th}^2 = \frac{3}{2}kT$$

The effective mass of electrons in silicon is $0.26m_0$. According to the thermal velocity formula, the thermal velocity of silicon at 300K is about $2.3 \times 10^5 m/s$. According to $q\tau = m_e^*\mu$, the average free time $\tau \approx 2.96 \times 10^{-13} s$. according to the thermal velocity formula, $l = V_{th} \cdot \tau \approx 68\ nm$ is estimated, which is approximately consistent with the calculation results of the above formula.

## D. Avalanche excess noise

White noise in APD includes avalanche noise (excessive noise) in addition to thermal noise or Johnson noise and shot noise. Generally, before avalanche breakdown, the current noise level shows similar intensity with the change of voltage, but it will increase rapidly with the multiplication factor. The excess noise of APD can be calculated by the following formula,

$$S(f) = 2eI_{ug}M^2F(M)$$

$$M = (I_{ph} - I_{dark})/I_{ug}$$

$$F(M) = M\left[1 - \left(\frac{1-k}{k}\right)\left(\frac{M-1}{M}\right)^2\right]$$

Where, $I_{ug}$ is the photocurrent at $M = 1$, $M$ is the current multiplication factor, $F(M)$ is the excess noise factor, $k$ is the ionization coefficient, $k = \alpha_h/\alpha_e$. Theoretically, the limit of excessive noise is the noise when $F(M) = 1$. At this time, the noise power spectral density is $S(f)/I^2 = 2eF(M)/I_{ug}$.

## E. Noise limit and signal-to-noise ratio

For the limit estimation of thermal noise, taking graphene/Silicon Schottky junction as an example, under thermal equilibrium (V=0), according to the theory of hot electron emission,

$$I = I_s(\exp \frac{qV}{\eta kT} - 1)$$

$$I_s = AA^*T^2 \exp(\frac{-e\phi_B}{kT})$$

Where, $I$ is dark current, reverse saturation current $I_s = J \times A = 35.7\mu A/cm^2 \times 1 \times 10^{-4} cm^2 = 3.57 \times 10^{-9} A$, $J$ is the current density, $V$ is the Schottky junction voltage, $\eta$ is the ideal factor, $A$ is the Schottky junction contact area, $A^*$ is the effective Richardson constant of silicon, and $\phi_B$ is the Schottky barrier. $exp\left(\frac{qV}{\eta kT}\right) \approx 1 + \frac{qV}{\eta kT}$, $\eta \approx 1.203$, According to the formula, $I \approx I_s(\frac{qV}{\eta kT})$, The junction resistance $R_i$ under thermal balance is the partial derivative of $V$, $R_i = \frac{dV}{dI}|_{V=0} = \frac{\eta kT}{I_s \cdot q} \approx 8.71 \times 10^6 \Omega$. From the spectral density formula of thermal noise, $S_I(f) = 4k_BT/R_i \approx 1.9 \times 10^{-27} A^2/Hz$. The spectral density of shot noise is given by Schottky theorem, $S_I(f) = 2q<I> \approx 1.14 \times 10^{-27} A^2/Hz$. The total noise spectral density of the device is $S_{ALL}(f) = \frac{4k_BT}{R_i} + 2q<I> + I^2 \cdot \frac{\alpha_H}{Nf}$. The total noise current $I_{ALL}$ is the integral of the total spectral density within the measurement frequency range $[f_1, f_2]$.

Theoretically, the randomness of photon emission conforms to Poisson distribution, which is more significant in weak light. In unit time, the probability of $n$ photons emitted by the light source meets, $p\{x = n\} = \frac{\mu^n}{n!} \cdot e^{-\mu}$, number of photogenerated carriers $m = \eta \times n$. Ideally, the signal-to-noise ratio of the device can be expressed as $SNR = \langle m \rangle^2 / \sigma_m^2 = \langle m \rangle$, $\langle m \rangle$ is the expected number of photogenerated carriers, $\sigma_m^2$ is the fluctuation variance of the number of photo carriers. Considering that the actual device operates with large avalanche gain, the circuit noise includes several parts: (1) thermal noise or Johnson noise; (2) Shot noise; (3) Avalanche gain noise; (4) 1/f noise (usually occurs when f<100khz). The signal-to-noise ratio formula is modified as, $SNR = \frac{\langle m \rangle^2}{\langle m \rangle F + \sigma_q^2/\langle G \rangle^2}$, where $\sigma_q$ is the circuit noise, gain $G = \frac{1.24 \cdot R}{\alpha \cdot \lambda(\mu m)}$, $F$ is the excess gain noise and $R$ is the responsivity. Where $\sigma_q^2 = \frac{I_{ALL}}{4e^2 \cdot \Delta f^2}$.

## Supplementary Section 2

### The Carrier Multiplication Caculations

The specific efficiency of the CM can be described by the following formula:

$$CM(t) \equiv \frac{n_c(t) - n_c(-\infty)}{n_c(0) - n_c(-\infty)}$$

Among them $n_c(t) = \sum_\mu \int_0^\infty d\varepsilon f_\mu(\varepsilon) v(\varepsilon)$ The number of electrons per unit cell in the conduction band at time t. $v(\varepsilon) = \frac{2\varepsilon A_0}{2\pi(\hbar v)^2}$ This is the density of states under MDF (Maxwell Distribution Function in equilibrium). Among them, v represents the Fermi velocity.

After moiré graphene is optically excited, following a non-equilibrium distribution, it undergoes a relaxation phase. During the relaxation phase, hot electrons transfer heat to the lattice, a process we refer to as the cooling process of moiré graphene. There are multiple cooling mechanisms in moiré graphene, primarily including the following four methods.

Including Normal collision cooling (which encompasses intrinsic optical phonon dissipation modes and out-of-plane dissipation modes) and Wiedemann-Franz cooling.

For different heat dissipation modes, different models are provided:

*Normal collision cooling*

Below, we describe a conventional cooling method, with the lattice temperature denoted as $T_L$. The electron density is $n$, and the entire cooling process is characterized by the temporal evolution of the electron temperature $T_e(t)$. The electron temperature $T_e$ is determined by the chemical potential $\mu$, while $n$ is determined by the distribution function. When electron-electron (e-e) scattering concludes (in this modeling, it is assumed that only e-e scattering contributes to heating the modeled system) and the system reaches thermal equilibrium, the system begins to cool. At this point, when the electron temperature equals the lattice temperature, the cooling process is complete. Therefore, the variation in electron temperature can be expressed as:

$$T_e(t) = \frac{T_0}{\sqrt{t/\tau_0 + 1}}$$

The characteristic time $\tau_0$

$$\tau_0 = \frac{424}{D^2 T_0^2} \ \mu s.$$

$D$ represents the deformation potential. Generally, data obtained through experiments are substituted into the above formula to obtain the deformation potential $D$. Using the Boltzmann equation, we can obtain:

$$Q = \partial_t \sum_{\mathbf{k}\alpha} \epsilon_{k\alpha} f_{\mathbf{k}}^{\alpha} = \sum_{\mathbf{k}\alpha} \epsilon_{k\alpha} S_{ph}(f_{\mathbf{k}}^{\alpha})$$

Describes the cooling rate caused by intrinsic optical phonon modes:

$$\mathcal{Q}_o = \frac{g^2 \omega_0^4}{(2\pi v_g^2)^2} [N_e(\omega_0) - N_L(\omega_0)] \mathcal{F}(T_e, \mu)$$

$$\mathcal{F}(T_e, \mu) = \int_{-\infty}^{\infty} dx |x(x-1)|[f([x-1]\omega_0) - f(x\omega_0)]$$

$N_e$ and $N_L$ correspond to the Bose distributions at temperatures $T_e$ and $T_L$, respectively.

*Wiedemann-Franz cooling*

$$dT_{el}/dt = -\gamma_1(T_{el} - T_0)$$

In this, $\gamma_1$ denotes the cooling rate.

$$\gamma_1 = \frac{3V^2\mu^3}{4\pi^2\hbar^3\rho v_F^4 k_B T_{el}} \approx 0.87 \frac{(\mu[\text{meV}]/100)^3}{T_{el}[\text{K}]/300} \text{ns}^{-1}$$

The following is a specific description of carrier dynamics in moiré graphene.

$V$ is the deformation potential constant, $\mu$ is the Fermi level, and $\rho$ is the surface density of moiré graphene. This formula indicates that the cooling rate mediated by acoustic phonons in moiré graphene strongly depends on temperature and the Fermi level. The larger the Fermi level $\mu$ (i.e., the higher the doping concentration), the more significant the cooling rate $\gamma$, which is proportional to $\mu^3$. Conversely, the higher the electron temperature $T_{el}$, the lower the cooling rate, inversely proportional to $T_{el}$. The carrier dynamics in moiré graphene can be described by the Bloch equations of moiré graphene:

The input light is described using the vector A(t).

$p_k$ It is a measure of the probability of transition.。

$$p_k(t) = \langle a_{vk}^\dagger a_{ck} \rangle(t)$$

$\rho_k^\lambda$ This represents that optical excitation changes the occupation probabilities in the two bands. Here, λ = v, c denotes the valence band and the conduction band, respectively.

$$\rho_k^\dagger(t) = \langle a_{\lambda k}^\dagger a_{\lambda k} \rangle(t)$$

$n_q^j$ represents the probability of an increase in phonon occupation following the relaxation of excited charge carriers. Here, j denotes the different phonon modes.($j = \Gamma\text{-LO}, \Gamma\text{-TO}, K, \Gamma\text{-LA}$)

$$n_q^j(t) = \langle b_{jq}^\dagger b_{jq}\rangle(t)$$

Next, we need to determine the time evolution of these three quantities. The Bloch equations for moiré graphene are:

$$\dot{p}_{\boldsymbol{k}} = (i\Delta\omega_{\boldsymbol{k}} + \Omega_{\boldsymbol{k}}^{\lambda\lambda})p_{\boldsymbol{k}} - i\Omega_{\boldsymbol{k}}^{vc}(\rho_{\boldsymbol{k}}^c - \rho_{\boldsymbol{k}}^v) + \dot{p}_{\boldsymbol{k}}|_{\text{hf+s}}$$

$$\dot{\rho}_{\boldsymbol{k}}^v = -2\Im m(\Omega_{\boldsymbol{k}}^* p_{\boldsymbol{k}}) + \dot{\rho}_{\boldsymbol{k}}^v|_{\text{hf+s}}$$

$$\dot{n}_{\boldsymbol{q}}^j = -\gamma_j(n_{\boldsymbol{q}}^j - n_B) + \dot{n}_{\boldsymbol{q}}^j\big|_s$$

The energy gap is:

$$\hbar\Delta\omega_{\boldsymbol{k}} = (\varepsilon_{\boldsymbol{k}}^v - \varepsilon_{\boldsymbol{k}}^c)$$

The Rabi frequency is::

$$\Omega_{\boldsymbol{k}}^{vc}(t) = i\frac{e_0}{m_0}\boldsymbol{M}(\boldsymbol{k})^{vc}\cdot \boldsymbol{A}(t)$$

The lifetime of the phenomenological phonon is:

$$\gamma_j^{-1} = 1.2\text{ps}$$

The $\rho_{\boldsymbol{k}}^\lambda$ and $n_{\boldsymbol{q}}^j$ distributions are described here, representing two processes that can describe the doubling process and the cooling process mentioned earlier, respectively.

In the Bloch equation of moiré graphene, the many-body interactions are divided into the Hartree-Fock term $\dot{p}_{\boldsymbol{k}}|_{\text{hf}}$ and the scattering term $\dot{p}_{\boldsymbol{k}}|_s$. Carrier multiplication focuses on the scattering probability. To describe the scattering probability, one should start from the coherence of the particles, as interactions among multiple particles lead to decoherence processes. The diagonal dephasing rate can be further described.

$$\gamma_{2,\boldsymbol{k}}(t) = \frac{1}{2}\sum_\lambda [\Gamma_{\lambda,\boldsymbol{k}}^{\text{in}}(t) + \Gamma_{\lambda,\boldsymbol{k}}^{\text{out}}(t)]$$

Non-diagonal phase cancellation is further described as

$$\mathcal{U}_{\boldsymbol{k}}(t) = \sum_{\boldsymbol{k}'}[T_{\boldsymbol{k},\boldsymbol{k}'}^a(t)p_{\boldsymbol{k}'}(t) + T_{\boldsymbol{k},\boldsymbol{k}'}^b(t)p_{\boldsymbol{k}'}^*(t)]$$

In this work, we consider only the off-diagonal dephasing under Coulomb interactions:

$$T^i_{\mathbf{k},\mathbf{k}'} = \frac{\pi}{\hbar} \sum_{l_1,l_2,\lambda} [\hat{V}^{\mathbf{k}c,l_2}_{\mathbf{k}'\lambda_i,l_1} \hat{V}^{\mathbf{k}'\lambda'_i,l_1}_{\mathbf{k}v,l_2} \mathcal{T}^{\mathbf{k}\lambda}_{l_1,l_2} \delta(\varepsilon^\lambda_\mathbf{k} \mp \varepsilon^\lambda_{\mathbf{k}'} - \varepsilon_{l_1} + \varepsilon_{l_2}) + V^{\mathbf{k}c,\mathbf{k}'\lambda'_i}_{l_2,l_3} \hat{V}^{l_2,l_3}_{\mathbf{k}v,\mathbf{k}'\lambda_i} \tilde{\mathcal{T}}^{\mathbf{k}\lambda}_{l_1,l_2} \delta(\varepsilon^\lambda_\mathbf{k} \mp \varepsilon^\lambda_{\mathbf{k}'}$$
$$- \varepsilon_{l_1} - \varepsilon_{l_2})]$$

Among them $\lambda_i = c, \lambda'_i = v (\lambda_i = v, \lambda'_i = c)$

$$\mathcal{T}^{\mathbf{k}\lambda}_{l_1,l_2} = (1-\rho_{l_1})\rho_{l_2}\rho^\lambda_\mathbf{k} + \rho_{l_1}(1-\rho_{l_2})(1-\rho^\lambda_\mathbf{k})$$

$$\tilde{\mathcal{T}}^{\mathbf{k}\lambda}_{l_1,l_2} = (1-\rho^\lambda_\mathbf{k})\rho_{l_1}\rho_{l_2} + \rho^\lambda_\mathbf{k}(1-\rho_{l_1})(1-\rho_{l_2})$$

$$\hat{V}^{l_1,l_2}_{l_3,l_4} \equiv V^{l_1,l_2}_{l_3,l_4} - V^{l_2,l_1}_{l_3,l_4}$$

Among them, the scattering probability related to time and momentum:

$$\Gamma^{in/out}_l(t) = \Gamma^{in/out}_{l,cc}(t) + \Gamma^{in/out}_{l,cp}(t)$$

It is divided into the carrier-carrier scattering part and the carrier-phonon scattering part. In the case of only Coulomb scattering, we present the formula for the carrier-carrier scattering part:

$$\Gamma^{in/out}_{l,cc}(t) = \frac{2\pi}{\hbar} \sum_{l_1,l_2,l_3} V^{l,l_1}_{l_2,l_3} \left(2V^{l,l_1*}_{l_2,l_3} - V^{l,l_1*}_{l_3,l_2}\right)$$
$$\times \mathcal{R}^{in/out,cc}(t) \delta(\varepsilon_1 + \varepsilon_{l_1} - \varepsilon_{l_2} - \varepsilon_{l_3})$$

Among them

$$\mathcal{R}^{in,cc}(t) = [1-\rho_{l_1}(t)]\rho_{l_2}(t)\rho_{l_3}(t)$$

$$\mathcal{R}^{out,cc}(t) = \rho_{l_1}(t)[1-\rho_{l_2}(t)][1-\rho_{l_3}(t)]$$

These two items clearly include the obstruction term in the bubble.

The efficiency of the scattering channel is determined by the Coulomb matrix element $V^{l,l_1}_{l_2,l_3}$

The Coulomb matrix elements here contain all the matrix elements of the Coulomb interaction, including the matrix elements of Auger composite and collision ionization.

# Supplementary Section 3

## The Density of States Calculations

We use the theoretical formula:

$$\cos(\theta) = \frac{n^2 + 4nm + m^2}{2(n^2 + nm + m^2)}$$

The **moiré** graphene with 5 layers of corner corners is modeled at 10° each, and the **moiré** graphene with 5 layers with a turning angle of 10° is modeled by Spyder software. We imported the ASE library for modeling, the idea of modeling is to calculate the values of m and n first, we get the data as 19 and 14. So we get graphene with a angle of 10° in two layers, and the next layer is also the same m and n, but there will be an overlap effect with the corner of the second layer, so we add the compensation angle on the basis to achieve the subsequent number of layers at a larger angle. Finally, output the vasp file.

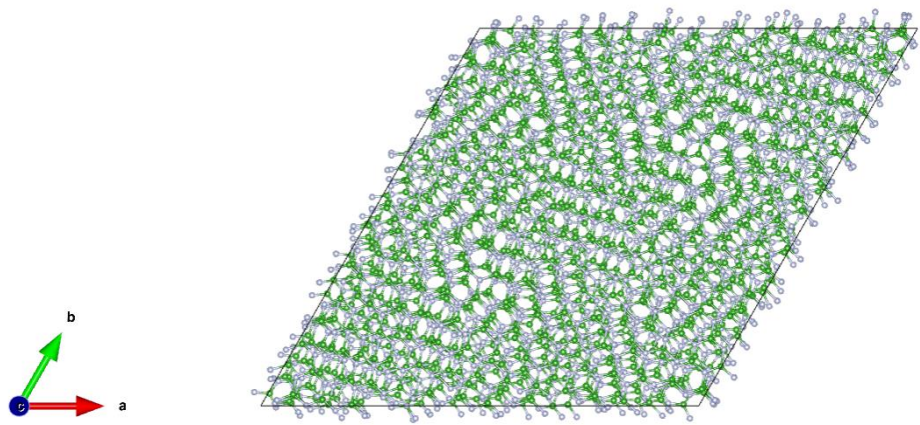

Since we are outputting 5 files with different layers, we need to write a .sh. The script implements stacking operations on each layer. Finally, output a stacked file, and the vasp file can be visualized through the Vesta software.

After we get the file, we set the file as a POSCAR file, and calculate the electronic state density of the moiré graphene system with a corner of 10° in 5 layers, based on the Vienna Ab initio Simulation Package (VASP) platform, we need to integrate the POSCAR, INCAR, POTCAR, and KPOINTS files into one folder. The INCAR file is based on the density functional theory (DFT) calculation requirements to configure key parameters to ensure the accuracy and reliability of the state density calculation, PThe OTCAR file is a pseudo-file of the C element. The KPOINTS file uses the Monkhorst-Pack(MP) k-point grid for Brillouin Zone (BZ) sampling to ensure the convergence of the Brillouin Zone integral through high-density k-point sampling to avoid the statistical bias of state density caused by k-point sparseness.

After modifying the above folder, we log in to the supercomputing platform, and the node we use here is configured to 96 cores, memory size is 1.5TB single node. On this node, we use a Linux-based cluster architecture to transmit the above working directory via SCP/SFTP protocolto the specified storage path to the supercomputing node. Write the job submission script in the above folder. Using VASP as the core, the density functional theory is used to focus on "Electronic Structure Solving-Brillouin Area Statistics-State Density Generation". Its core equation formula is the Kohn-Sham equation:

$$[-\frac{\hbar^2}{2m_e}\nabla^2 + V_{eff}(r)]\psi_{s,k}(r) = E_{s,k}\psi_{s,k}(r)$$

$\psi_{s,k}(r)$ The spin is s and the wave vector is k, the $V_{eff}(r)$, effective potential field, and the $E_{s,k}$ single electron energy level. VASP calculation of DOS needs to be divided into two steps: "self-consistent calculation (SCF)" and "non-self-consistent calculation (NSCF)", the principle of which is to ensure the accuracy of the electronic structure first, and then ensure the reliability of state density statistics, both of which are indispensable.

Self-consistent computation (SCF) - The core purpose of obtaining convergent electronic states (wave functions and energy levels) SCF is to solve the convergent single-electron wave function $\psi_{s,k}(r)$ and energy level $E_{s,k}$, because DOS is right $E_{s,k}$ If the electronic state itself is inaccurate, the DOS results are meaningless. Step 2: Non-Self-Consistent Calculation (NSCF) - Dense Sampling Statistics The core purpose of NSCF is to perform dense K-point sampling of the Brillouin region, calculate the energy distribution of all electronic states, and obtain a smooth and accurate DOS.

Finally, we can visualize the output file in Vesta to obtain a state density image.

# Supplementary Section 4

# Thermalized Optical Phonon Effect

The thermophonon, the thermophonon we describe here is the "thermal optical phonon". The difference between optical phonons and general optical phonons here is that the general optical phonons are optical phonons in thermal equilibrium, and the number of phonons occupied by the macroscopic lattice temperature of the material is determined by the temperature of the macroscopic crystal lattice of the material, following the Boltzmann distribution. The thermo-optical phonons are in a non-equilibrium state, and the number of phonons occupies far more than the expected thermal equilibrium, and the corresponding effective temperature is much higher than the lattice temperature.

In general optical phonons and thermo-optical phonons, we use the Bose-Einstein distribution, both of which belong to the boson and do not follow the bubble blocking effect. We generally describe it with the Bose-Einstein distribution, and the general optical phonons here refer to the optical phonons in thermal equilibrium, and their occupancy is completely determined by the lattice macroscopic thermal equilibrium temperature. The formula for distribution is:

$$n(\omega) = \frac{1}{e^{\hbar\omega/(k_B T_{lattice})} - 1}$$

At this time, $T_{lattice}$ the lattice temperature is the crystal lattice temperature. But the Bose-Einstein distribution in quasi-thermal equilibrium is:

$$n(\omega) = \frac{1}{e^{\hbar\omega/(k_B T_{\text{phon}})} - 1}$$

The hot phonon effect we mentioned here is that the energy we give in an instant and its high level cause the optical phonons to gather at this time and cannot dissipate heat, resulting in an optical phonon bottleneck.

Thermophonons are generated because the rate of energy generation is higher than the rate of energy dissipation.

Supplement two bottlenecks:

1. Thermo-optic phonon bottleneck:

If the non-harmonic scattering (decay) efficiency of optical phonons is extremely low (e.g., it is difficult to quickly decay into acoustic phonons and transfer energy to the substrate), thermo-optical phonons accumulate in large quantities, resulting in "blocked" energy exchange between carriers and optical phonons - the carriers are difficult to cool further by "emitting optical phonons" (because the optical phonons are already "hot" and it becomes difficult to receive the energy of the carriers).

The "heat accumulation" and "slow dissipation" of the optical phonons themselves hinder the cooling of the carriers.

2. Optical phonon energy bottleneck:

The energy of the optical phonon is quantized (with a minimum energy threshold, such as the optical phonon energy ~200 meV in moiré graphene). When the carrier energy is reduced to close to or below the optical phonon energy, the process of losing energy by the carrier by "emitting optical phonons" becomes extremely difficult (because the energy is not enough to trigger a full optical phonon emission). At this point, carrier cooling can only rely on slower processes (such as scattering with acoustic phonons), resulting in a sudden drop in cooling rate.

The carrier's own energy is too low to meet the energy matching requirements of "emitting optical phonons", and the optical phonon energy becomes the "threshold" for cooling.

The above two bottlenecks are compared to the competitive relationship between the hot phonon effect and the supercollision effect. I think the hot phonon effect mentioned in our project leans more towards the first one.

At low temperatures: Due to the proximity of the carrier energy distribution to the Dirac store, the efficiency of optical phonon emission drops sharply The "thermo-optical phonons" that have been generated are difficult to dissipate through non-simple harmonic scattering (decay into acoustic phonons), forming an "energy bottleneck" that causes the subsequent cooling of the carriers (dependent on acoustic phonons or optical phonon decay) becoming extremely slow.

Core formula:

$$C_e(T_e(t))\partial_t T_e(t) = -\hbar\sum_\alpha \omega_\alpha \mathcal{R}_\alpha(T_e(t), T_\alpha(t))$$

$$D(\omega_\alpha, T_\alpha(t))\partial_t T_\alpha(t) = \frac{\mathcal{R}_\alpha(T_e(t), T_\alpha(t))}{M_\alpha(T_e(t))} - \gamma_\alpha\left[n_\alpha(T_\alpha(t)) - n_\alpha\left(T_\alpha^{(0)}\right)\right]$$

$C_e(T_e(t))$ electron heat capacity (changes with electron temperature, reflecting the ability of electrons to store energy); $\partial_t T_e(t)$ derivative of electron temperature to time; $\alpha$ optical phonon pattern marker; $\omega_\alpha$: the angular frequency of the phonon in $\alpha$ mode; $\mathcal{R}_\alpha(T_e(t), T_\alpha(t))$: Emissivity of phonons in $\alpha$ mode (the number of phonons emitted by electrons per unit time); $D(\omega_\alpha, T_\alpha(t))$ phonons occupy the partial derivative of the number of pairs of temperature; $M_\alpha(T_e(t))$ the state density of the heated region in the phonon Brillouin region (varies with $T_e$ and is related to the electron-phonon momentum exchange range); $\gamma_\alpha$ amping rate of optical phonon relaxation to acoustic phonon; $n_\alpha(T)$ Bose-Einstein phonon occupancy function; $T_\alpha^{(0)}$ Lattice equilibrium temperature.

The first equation: electrons release energy by emitting optical phonons, causing the temperature of electrons to drop;

The second formula: the optical phonon (hot phonon) is heated by electrons on the one hand, and on the other hand, it is relaxed to the acoustic phonon through non-simple coupling relaxation, reflecting the "thermophonon bottleneck effect" (when the phonon generation rate > the relaxation rate, the hot phonon accumulation delays the cooling).

$$C_e(T_e) = \int_{-\infty}^{\infty} d\varepsilon \nu(|\varepsilon|)(\varepsilon - \mu)\left[-\frac{\partial f[(\varepsilon - \mu)/(k_B T_e)]}{\partial \varepsilon}\right]\left[\frac{\varepsilon - \mu}{T_e} + \frac{\partial \mu}{\partial T_e}\right]$$

is the size of the electron heat capacity and the density of the medium $\nu(|\varepsilon|) = \frac{N_f |\varepsilon|}{2\pi \hbar^2 v_F^2}$; $\mu$ is the chemical potential; $\varepsilon$ It is the energy of electrons.

$$R_\alpha(T_e, T_\alpha) = \frac{Q_\alpha(T_e, T_\alpha)}{\hbar \omega_\alpha}$$

Expressions for phonon emission rates are described.

$$C_e(T_e)\partial_t T_e = -\sum_\alpha Q_\alpha(T_e, T_\alpha)$$

The rate equation for the conservation of electron energy. The partial derivative of electron temperature to time, $\partial_t T_e$ which describes the rate of change of electron temperature with time;

$$Q(T_e, T_\alpha) = \hbar \omega_\alpha \frac{4\pi}{\hbar N_f} \int_\alpha^2 [n_\alpha(T_\alpha) - n_\alpha(T_e)] [\tilde{Q}(\mu, \omega_\alpha) - \tilde{Q}(\mu + \omega_\alpha, \omega_\alpha)]$$

Represents the energy transfer rate of electrons to $\alpha$-mode optical phonons; $N_f$ spin-valley simplicity degree; $\tilde{g}_\alpha$ normalized electron-optical phonon coupling strength; $\tilde{Q}$ Electron energy integration function.

$$\tilde{g}_\alpha = \beta_\Gamma \sqrt{\frac{\hbar}{2\rho_m \omega_\alpha}}$$

Represents the normalized electron-optical phonon coupling strength; $\beta_\Gamma$ Deformation potential coupling constant (describes the rate of change of electron energy with lattice deformation, reflecting the coupling strength of "electron-lattice deformation-phonon"); $\rho_m$ The areal mass density of graphene. $\beta_\Gamma = \frac{3}{2}\frac{\partial t}{\partial a}$

$$\tilde{Q}(\mu, \omega) = \int_{-\infty}^{+\infty} d\varepsilon \nu(\varepsilon)\nu(\varepsilon - \omega)\left[f\left(\frac{\varepsilon - \mu}{k_B T_e}\right) - \Theta(-\varepsilon)\right].$$

represents the integral function of electron energy; $f\left(\frac{\varepsilon - \mu}{k_B T_e}\right)$ represents the Fermi-Dirac distribution; $\Theta(-\varepsilon)$ Step function.

$$D(\omega_\alpha, T_\alpha) = \frac{\partial n_\alpha(T_\alpha)}{\partial T_\alpha}$$

The temperature derivative that represents the phonon occupancy number

$$M_\alpha(T_e) = \frac{N_\alpha}{4\pi}\left[\left(\frac{\varepsilon_{max}(T_e)}{\hbar v_F}\right)^2 - \left(\frac{\varepsilon_{min}(T_e)}{\hbar v_F}\right)^2\right]$$

represents the effective state density of optical phonons; $N_\alpha$: degeneracy of $\alpha$-mode optical phonons; $\varepsilon_{max}(T_e)$ In electron-phonon coupling, the maximum phonon energy exchanged by electrons; Electron-phonon coupling, the $\varepsilon_{min}(T_e)$ smallest phonon energy exchanged by electrons.

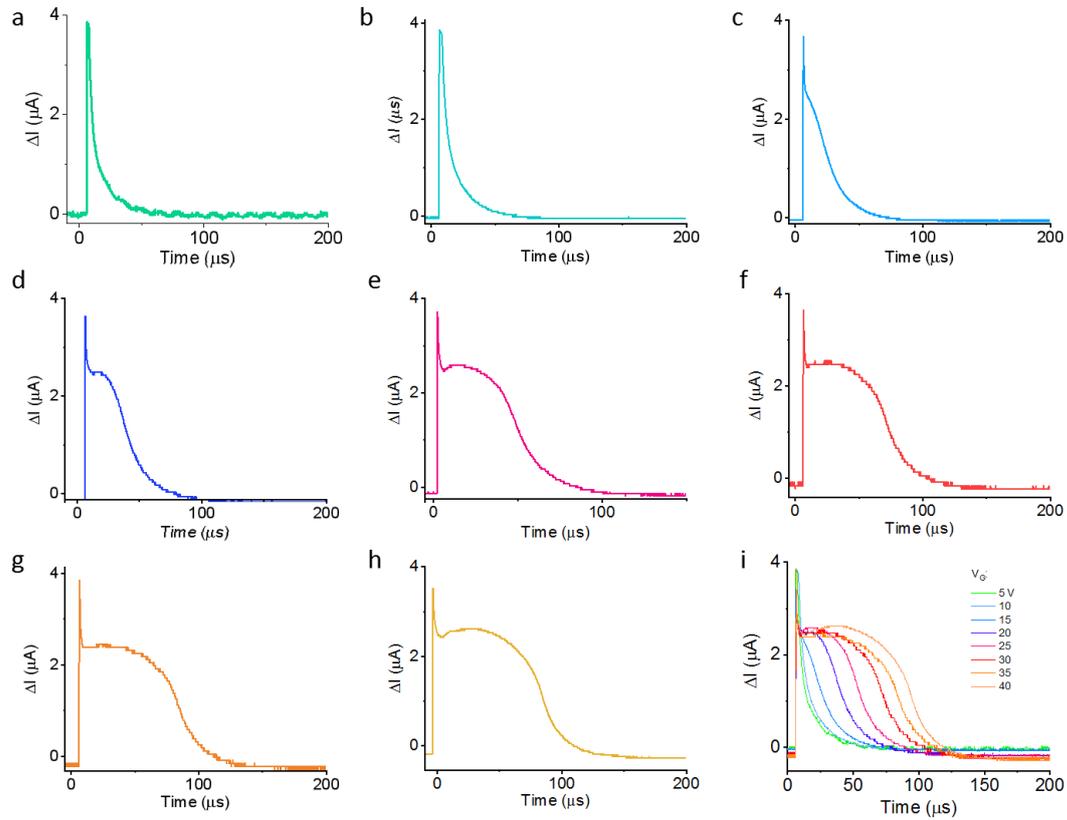

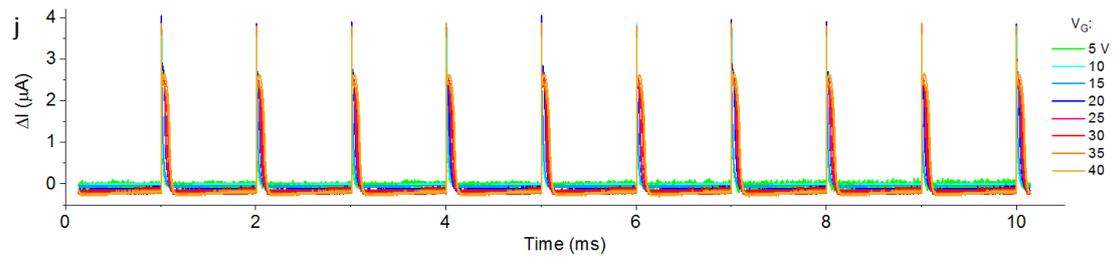

**SI Figure 1**. The photocurrent as a function of time under different gate voltages, where the laser wavelength is 1550 nm and pulsed width is 200 fs. The average laser power is about 100 nW.

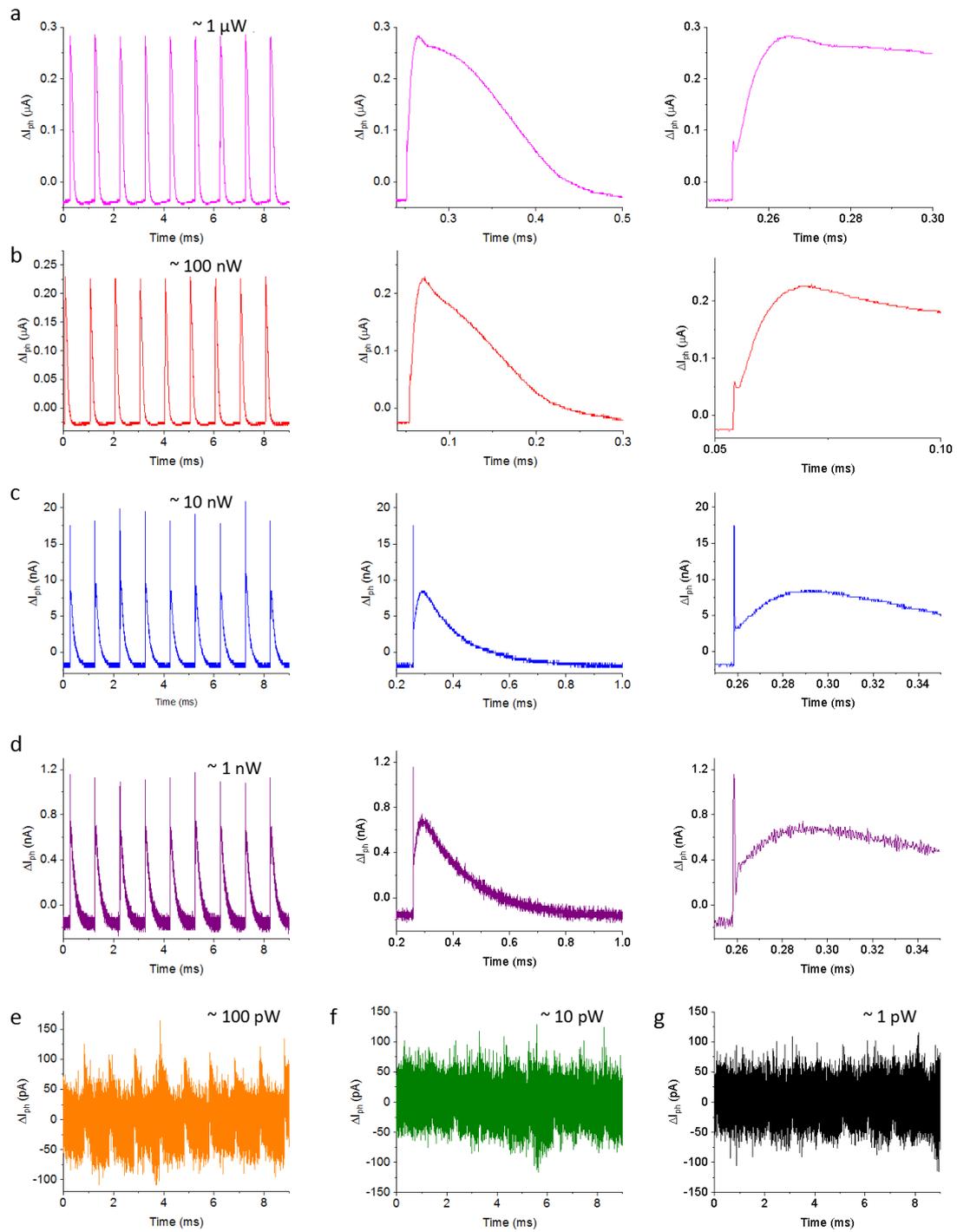

**SI Figure 2. a-g**, The photocurrent as a function of time under different laser power, where the laser wavelength is 1310 nm and pulse width is 200 fs.

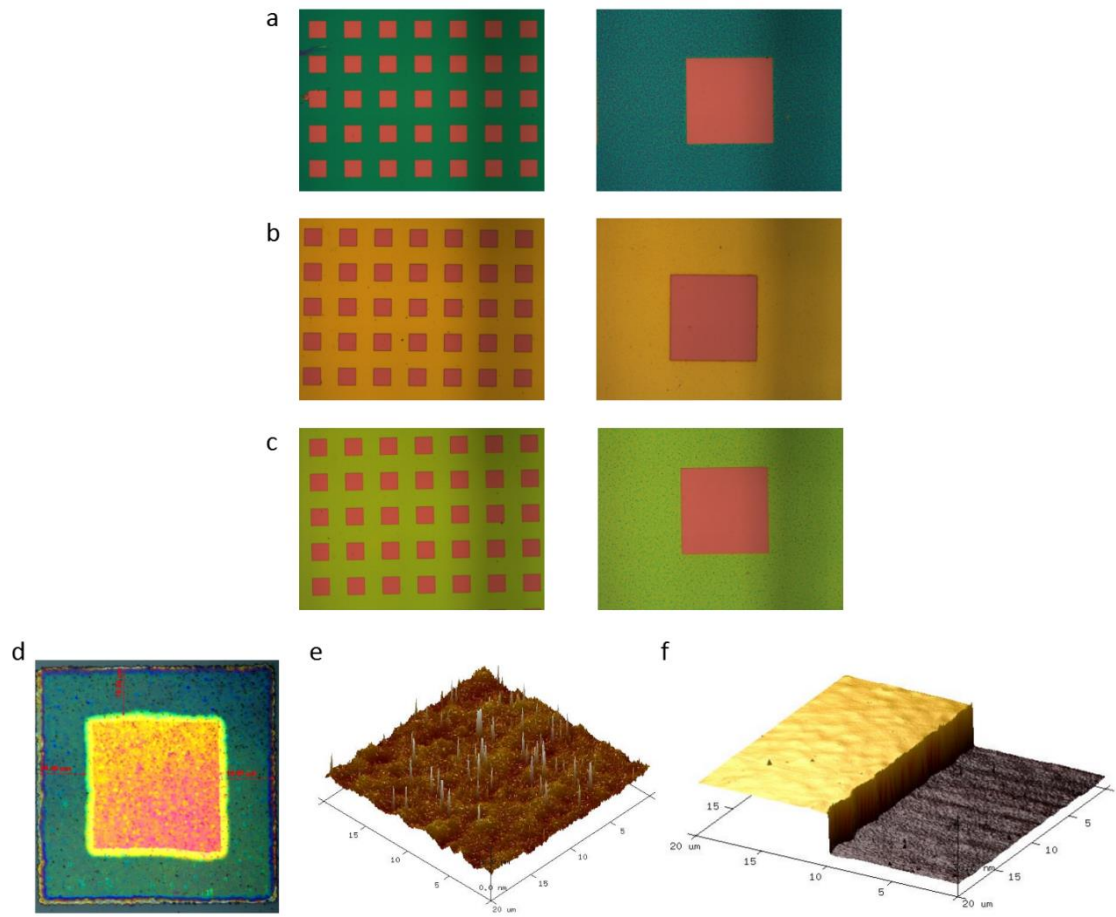

**SI Figure 3**. **The Fabricated Devices. a-c**, The light microscopic images of the etched SOI silicon with different thicknesses. **d**, The optical microscope top view of ICP etched silicon window. The AFM mappings of **e**, the etched silicon window and **f**, the step between silicon and silicon oxide.

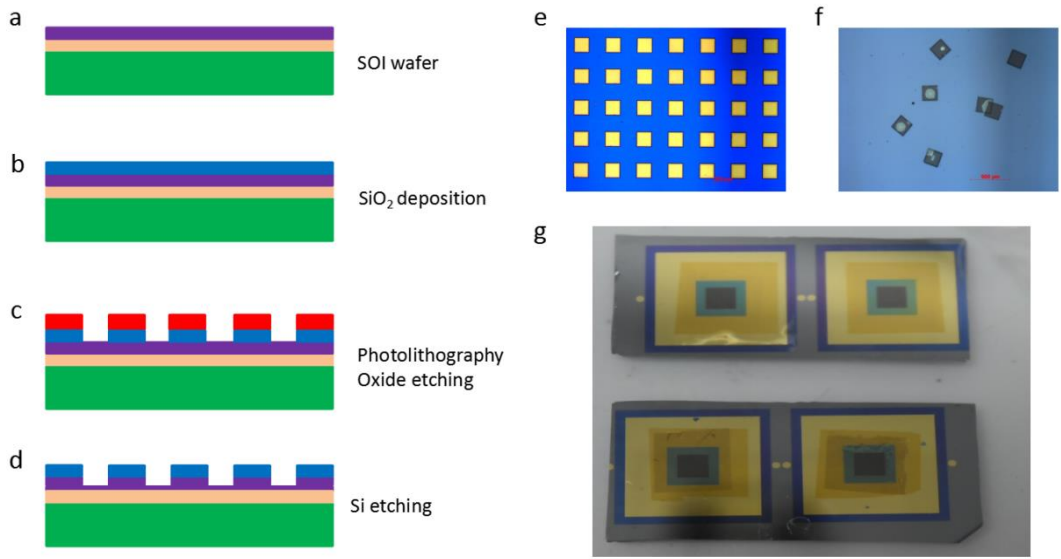

**SI Figure 4**. **a-d,** The schematic of the fabricated SOI silicon. **e-f,** The optical microscopic images of the etched SOI silicon. **g**, the fabricated SOI silicon windows and gold electrodes, where the moiré graphene layers on the top.

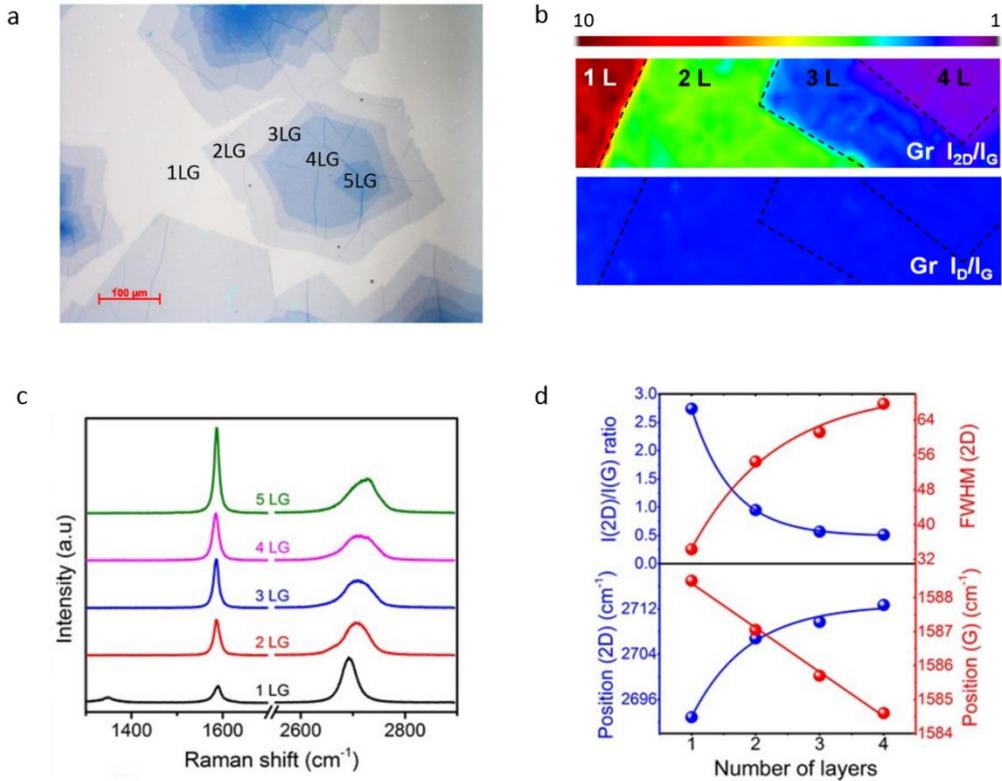

**SI Figure 5. a,** The optical image of the moiré graphene layers. **b,** The corresponding $I_{2D}/I_G$ and $I_D/I_G$ Raman mappings. **c,** The corresponding Raman spectra and **d,** their characterations.

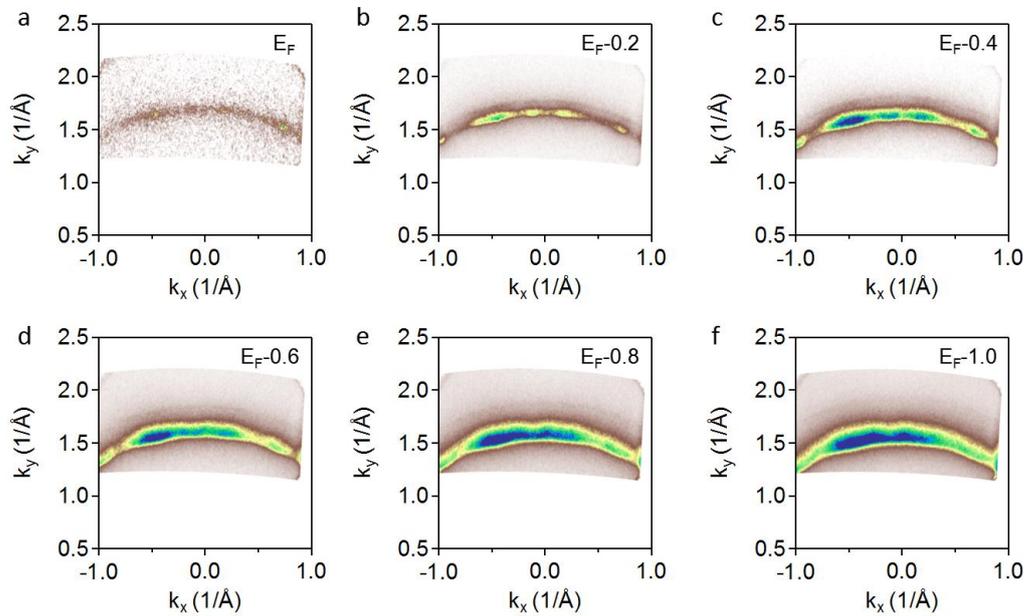

**SI Figure 6**. **The ARPES characterization of moiré graphene.** The two-dimensional curvature image of constant energy maps at energy **a**, $E_F$, **b**, $E_F - 0.2$ eV, **c**, $E_F - 0.4$ eV, **d**, $E_F - 0.6$ eV, **e**, $E_F - 0.8$ eV and **f**, $E_F - 1.0$ eV.

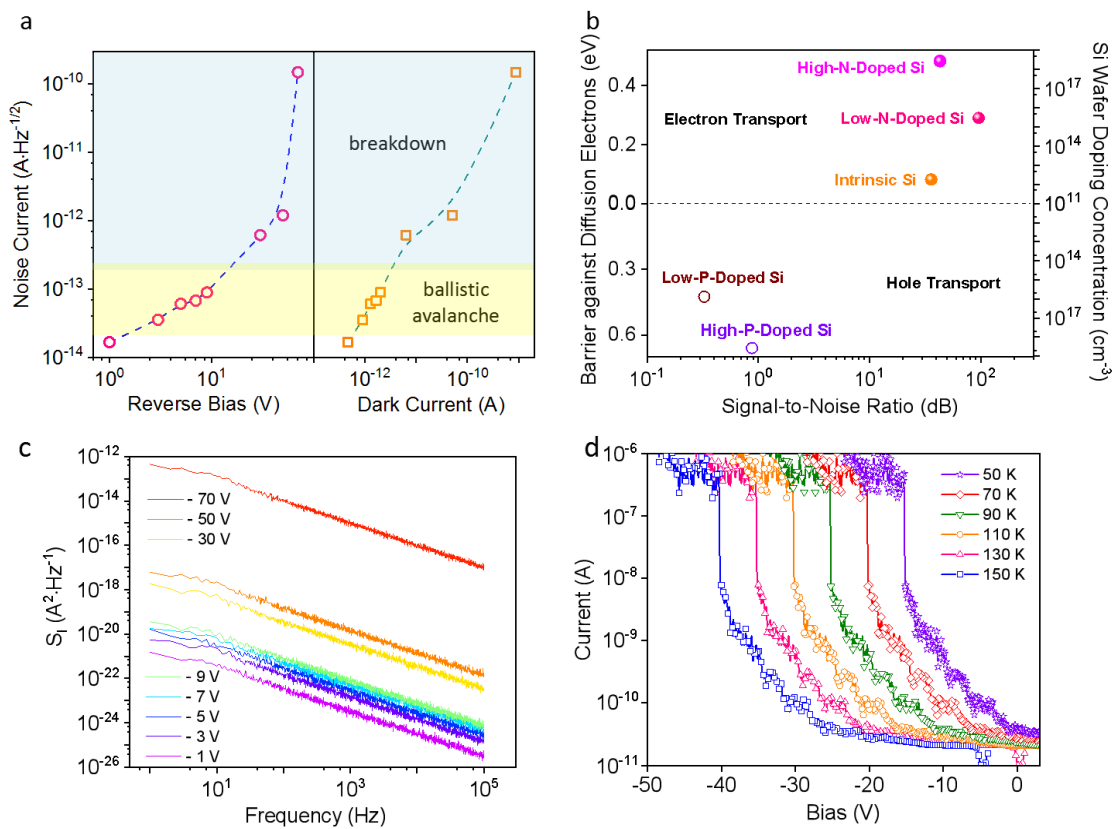

**SI Figure 7**. **Device Noise Characteristics**. **a**, the device noise current as a function of the bias voltage and dark current. **b**, Schottky barrier height and corresponding signal-to-noise ratio of silicon with different doping types and doping concentrations. **c**, The $S_I$ as a function of the scanning frequency at different bias voltages and measurement temperatures. **d**, the current-voltage curves of the device under different measurement temperatures.

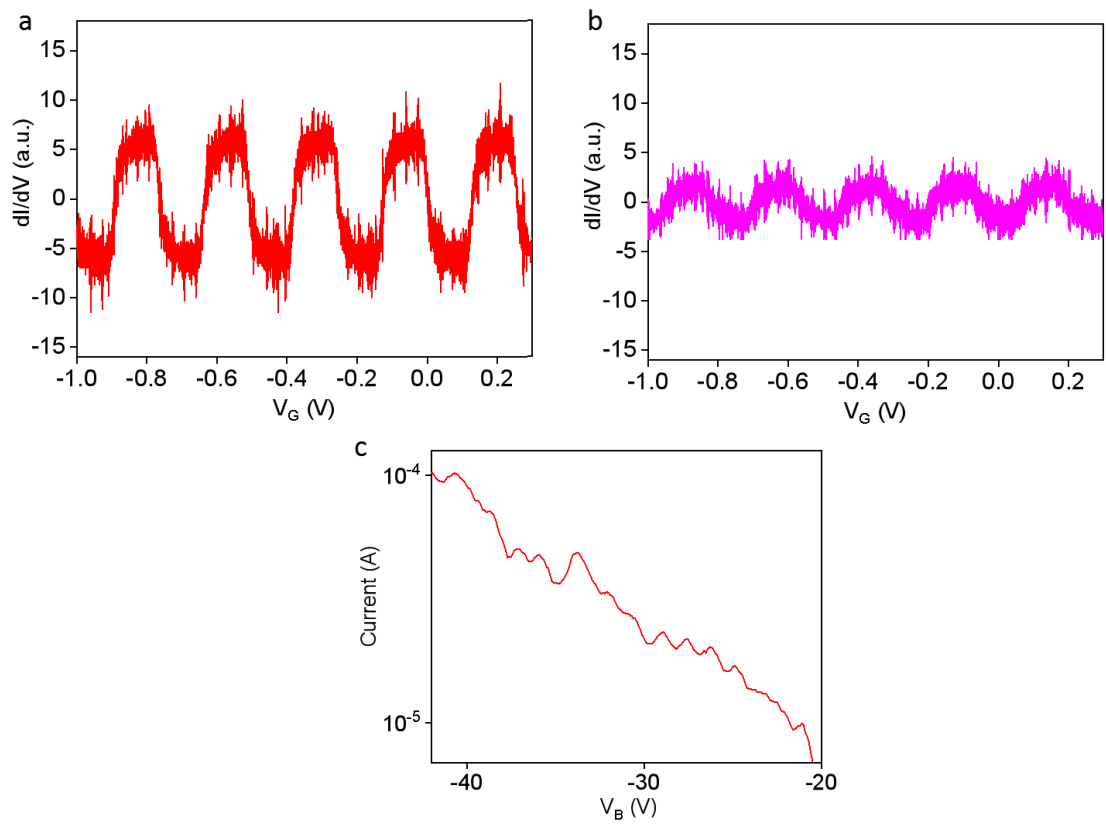

**SI Figure 8. a-b**, The dI/dV as a function of the gate voltage. **c**, The current as a function of the bias voltage.

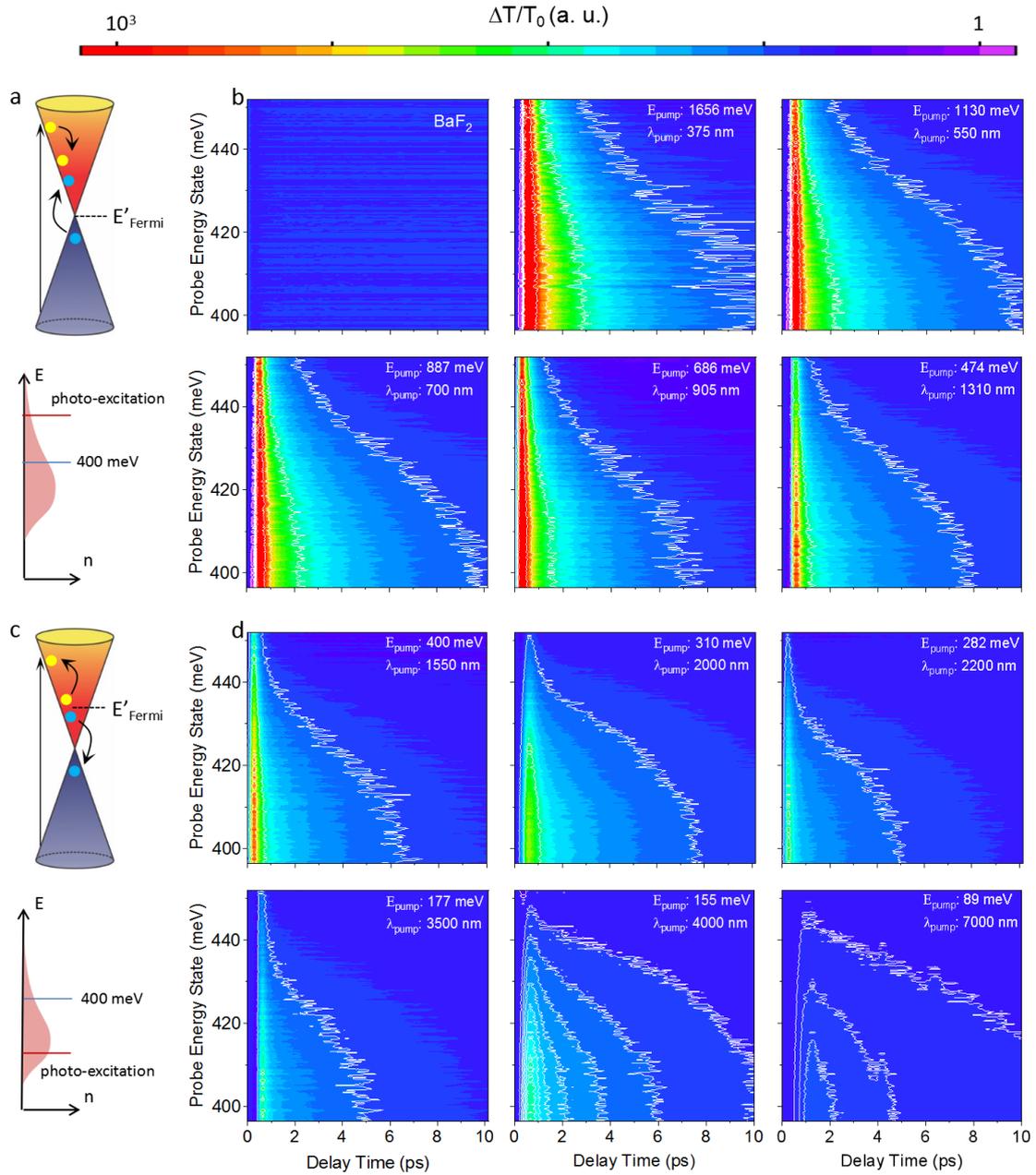

**SI Figure 9. Pump probe characterization of moiré graphene.** The schematic of the impact ionization induced **a,** carrier multiplication and **h,** Auger recombination with corresponding Fermi-Dirac distributions in a single layer graphene. The carrier density *n* as a function of the energy state *E* depicts the pump photo-excitation energy height and the bottom position of the probe enegy at 400 meV. The dynamical quasi Fermi level $E_{Fermi}$ is also shown to distinguish the inter- and intra- band transitions. The two-dimensional differential transmissivity countor maps as functions of the probe energy state and delay time of **b,** the $BaF_2$ substrate-only, and the graphene moiré superlattice on the $BaF_2$ substrate under pump photon energy $E_{pump}$ (and corresponding pump photon wavelength $\lambda_{pump}$) of **c,** 1656 meV (375 nm); **d,** 1130 meV (550 nm); **e,** 887 meV (700 nm); **f,** 686 meV (905 nm); **g,** 474 meV (1310 nm); **i,** 400 meV (1550

nm); **j**, 310 meV (2000 nm); **k**, 282 meV (2200 nm); **l**, 177 meV (3500 nm); **m**, 155 meV (4000 nm), **n**, 89 meV (7000 nm).

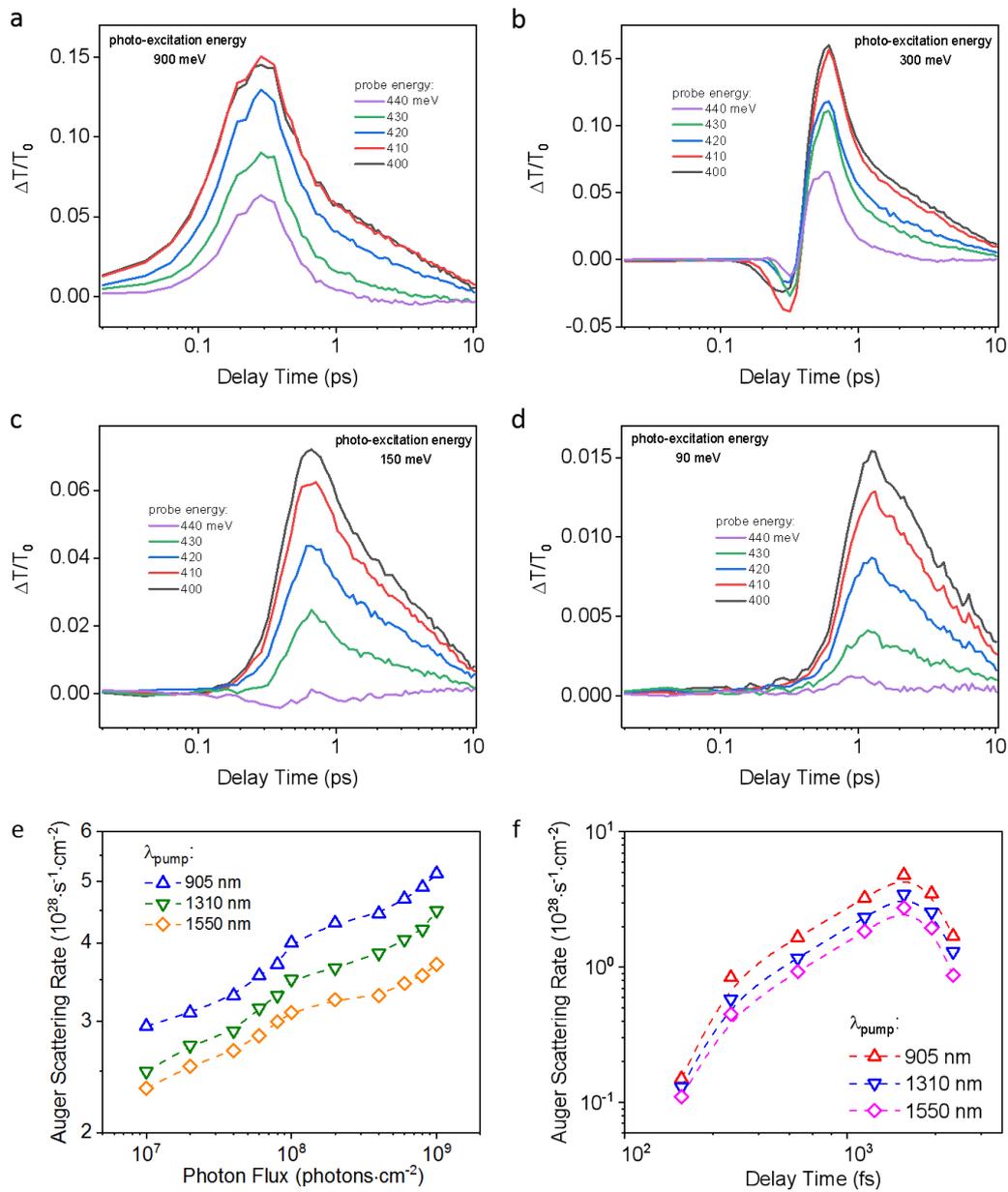

**SI Figure 10.** The differential transmission as a function of the probe laser energy, where the photo-excitation energies are **a,** 900 meV; **b**, 300 meV; **c,** 150 meV and **d,** 90 meV. The Auger scattering rate as a function of **e,** photon flux and **f,** delay time.

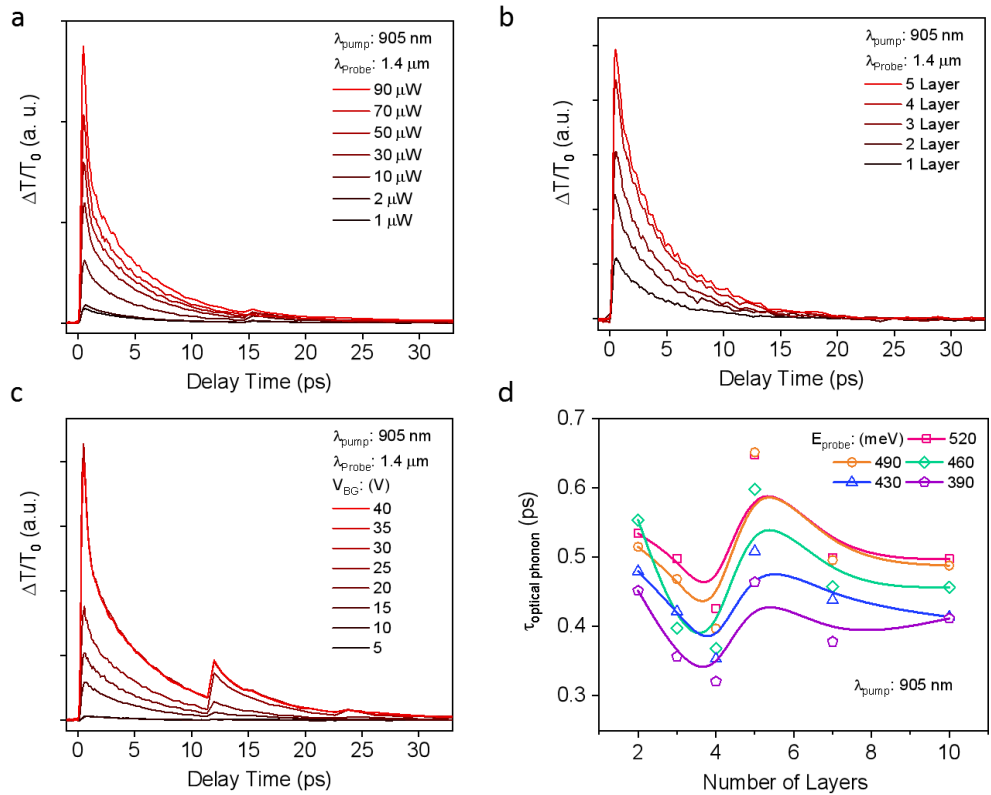

**SI Figure 11. a,** The differential transmission as a function of the pump laser power. **b**, The differential transmission as a function of the graphene layer number. **c, T**he differential transmission as a function of the back gate voltage. **d**, The lifetime of optical phonons as a function of the graphene layer number.

Table 1. The comparsion of the performance of relevant photodetectors in the literature.

| Materials | Wavelength(nm) | R (AW$^{-1}$) | D (Jones) | Gain | Optical power density (mW/cm²) | Response time | Temperature | Mechanism | Year | Ref. |
|---|---|---|---|---|---|---|---|---|---|---|
| SiO2-SiNWs/Gr | 1550 | 56.58 | / | 2.64×10$^4$ | / | 1.5us | Room temperature | Photoconductive | 2025 | 1 |
| Gr/InSe/Cr | 1550 | 0.067 | / | 1.8×10$^7$ | / | 2.9us | Room temperature | Avalanche | 2025 | 2 |
| InP+AlAs0.56Sb0.44+In0.53Ga0.47As | 1550 | 0.92 | / | 9 | / | / | Room temperature | Avalanche | 2019 | 3 |
| InP+GaAs0.5Sb0.5+Al0.85Ga0.15As0.56Sb0.44 | 1550 | 7418 | / | 278 | / | ~0.9ns | Room temperature | Avalanche | 2023 | 4 |
| Si MRR-PN diode | 1550 | 0.0728 | / | 72 | / | 45ps | Room temperature | Avalanche | 2013 | 5 |
| WS2/Bi2O2Se | 1550 | 0.021 | / | 12.88 | / | 410us | Room temperature | Avalanche | 2023 | 6 |
| Gr/si | 1550 | 0.00251 | / | 1123 | / | 1.4us | Room temperature | Avalanche | 2024 | 7 |
| QDS MCP | 1300 | 75.4 | / | 70 | / | / | Room temperature | Photocontrolled electric field regulation+ Charge tunneling - recycling effect | 2023 | 8 |
| Ge bipolar | 1550 | 2 | / | 7 | / | ~90ps | Room temperature | Photovoltaic | 2008 | 9 |
| Si/PtSi/TiO$_2$/Gr | 1310 | 9.1 | / | 4.7×10$^4$ | / | 300us | Room temperature | Photogating | 2024 | 10 |
| UCNPs/Gr/GaAs | 980 | 0.0176 | 1.1×10$^{11}$ | / | 19 | <0.04 s | Room temperature | Photovoltaic | 2018 | 11 |
| PdSe$_2$/Si | 780 | 0.3002 | 1.1×10$^{13}$ | / | 0.0009 | 38μs | Room temperature | Photovoltaic | 2018 | 12 |
| CdS/CIGS | 870 | 0.29 | 3.3×10$^{12}$ | / | 0.02 | <5 ms | Room temperature | Photovoltaic | 2022 | 13 |
| nMAG/epi-Si | 1550 | 0.00251 | 2.67×10$^9$ | / | 11000 | 1.4 μs | Room temperature | Avalanche | 2024 | 14 |
| Gr/SiO$_2$-SiNW-APD | 1550 | 56.58 | 8.64×10$^{12}$ | / | 67.3 | 1.5μs | Room temperature | Avalanche | 2025 | 15 |
| Gr/Ge | 1550 | 0.75 | 2.53×10$^9$ | / | 5.76 | <60 μs | Room temperature | Photovoltaic | 2019 | 16 |
| MoS$_2$/WSe$_2$ | 840 | 0.0022 | 2.4×10$^9$ | / | 0.066 | 46μs | Room temperature | Photovoltaic | 2025 | 17 |
| PtSe$_2$/InP | 940 | 0.718 | 4.37×10$^{12}$ | / | 0.5 | 4.35μs | Room temperature | Photovoltaic | 2024 | 18 |
| MoTe$_2$/Si | 980 | 0.33 | 2.9×10$^{11}$ | / | 0.0014 | 22μs | Room temperature | Photovoltaic | 2024 | 19 |